\documentclass[12pt,preprint]{aastex}
\shorttitle{tidal barrier}
\shortauthors{Li, Dobbs-Dixon, \& Lin}
\newcommand{\be}{\begin{equation}}
\newcommand{\ee}{\end{equation}}

\begin{document}

\title{Tidal Barrier and the Asymptotic Mass of Proto Gas-Giant Planets}
\author{Ian Dobbs-Dixon$^1$, Shu Lin Li$^{1,2}$, and D.~N.~C.~Lin$^{1}$}
\affil{$^1$Department of Astronomy and Astrophysics,
University of California, Santa Cruz, CA 95064, USA}
\affil{$^2$Department of Astronomy, Peking
University, Beijing, China}
 
\begin{abstract}
The extrasolar planets found with radial velocity surveys have masses
ranging from several Earth to several Jupiter masses. According to the
conventional sequential accretion scenario, these planets acquired
super-earth cores prior to the onset of efficient gas accretion with
rates that rapidly increase with their masses. In weak-line T-Tauri
disks, mass accretion onto protoplanetary cores may eventually be
quenched by a global depletion of gas, as in the case of Uranus and
Neptune. However, such a mechanism is unlikely to have stalled the
growth of some known planetary systems which contain relatively
low-mass and close-in planets along with more massive and longer
period companions.  Here, we suggest a potential solution for this
conundrum.  In general, supersonic infall of surrounding gas onto a
protoplanet is only possible interior to both of its Bondi and Roche
radii. At a critical mass, a protoplanet's Bondi and Roche radii are
equal to the disk thickness.  Above this mass, the protoplanets' tidal
perturbation induces the formation of a gap.  Although the disk gas
may continue to diffuse into the gap, the azimuthal flux across the
protoplanets' Roche lobe is quenched.  Using two different schemes, we
present the results of numerical simulations and analysis to show that
the accretion rate increases rapidly with the ratio of the
protoplanet's Roche to Bondi radii or equivalently to the disk
thickness. In regions with low geometric aspect ratios, gas accretion
is quenched with relatively low protoplanetary masses.  This effect is
important for determining the gas-giant planets' mass function, the
distribution of their masses within multiple planet systems around
solar type stars, and for suppressing the emergence of gas-giants
around low mass stars. Finally, we find that the accretion rate into
the protoplanets' Roche lobe declines gradually on a characteristic
timescale of a few Myr. During this final stage, the protracted
decline in the accretion flow across their Roche lobe and onto their
protoplanetary disks may lead to the formation and retention of their
regular satellites.
\end{abstract}


\section{Introduction}
In the conventional planet formation scenario \citep{safronov1969,
pollack1996,ida2004_1} heavy elements first condense into grains which
then grow into planetesimals. In a gaseous environment, the
hydrodynamic and tidal interaction between the disk gas and these
planetesimals damps their eccentricity \citep{Adachi1976,Ward1998}. 
Nevertheless, these planetesimals can perturb each other's orbit and
induce growth through coagulation \citep{aarseth1993,palmer1993}. This
growth process is stalled with the emergence of protoplanetary embryos
that have acquired all the residual planetesimals within their feeding
zone, which is 3-5 times that of their Roche radius
\citep{kokubo1998}. In protostellar disks with similar mass and
temperature distributions as those inferred for the primordial solar
nebula, the embryos attain isolation masses that are typically a few
times that of the Moon at 1AU \citep{lissauer1987,ida2004_1}.

In principle, the protoplanetary embryos can accrete gas from the disk
when they have attained sufficient mass such that their surface escape
speed is larger than the sound-speed in the disk. However, the gas
which falls onto the embryos also converts gravitational energy into
heat which is then transported to the surface of the nascent disk
\citep{Bodenheimer1986}. Unless the cores have acquired more than a
few earth masses ($M_\oplus$), the envelopes around them are too
tenuous to allow efficient radiative or convective transport
\citep{pollack1996}. Consequently, a quasi hydrostatic equilibrium is
established and gas sediments onto the cores on a protracted cooling
timescale.

Once the protoplanet acquires sufficient mass to allow gas to be
freely accreted, it enters a dynamical gas accretion phase.  During
this stage, the cores' gravity dominates the flow out to the Bondi
radius ($R_B$), where the local escape speed becomes comparable to the
gas sound-speed in the disk. This radius increases linearly with the
total mass of the protoplanet, which includes that of the core and the
envelope. The protoplanets' Roche radius ($R_R$) also increases with
the mass of the protoplanet, though the mass dependence is much
weaker. Although $R_B$ is generally smaller than $R_R$ for low-mass
protoplanets, $R_B$ overtakes $R_R$ for sufficiently large
protoplanetary masses.

As gas within $R_R$ collapses toward the cores, it is replenished by
the diffusion of gas from beyond $R_R$. The rate of matter diffusion
is generally assumed to be induced by turbulence, though the origin of
such turbulence in protostellar disks at a few AU's remains an
outstanding issue. Protoplanets also exert a tidal perturbation on the
disk with a strength that increases with their mass. Around relatively
massive protoplanets, this tidal perturbation is sufficiently strong
to induce the formation of a gap in the disk near the protoplanets'
orbits \citep{Goldreich1979,Goldreich1980} and to quench the accretion
\citep{lin1979,lin1985}. In relatively cold and weakly-viscous disks,
the asymptotic mass of the protoplanets near the ice line in the solar
nebula is comparable to that Jupiter \citep{lin1993,bryden1999}.

However, there are numerical simulations of disk-planet interaction
which indicate that gas continues to flow into the gap after its
formation conditions have been satisfied
\citep{Lubow1999,Kley2001,dangelo2002}. In these simulations, the
magnitude of both implicit and numerical viscosity is sufficiently
large that the intrinsic redistribution of angular momentum in the
disk gas is comparable to the tidal torque exerted by the embedded
protoplanet. Several simulations also take into account the effect of
the turbulent flow \citep{Armitage1998,Nelson2003,Laughlin2004}. All
these models show that gas continues to flow to the vicinity of the
protoplanet until its asymptotic mass becomes several times that of
the Jupiter.

Although radial velocity surveys have discovered many planets with
masses comparable to Saturn, the mass of Saturn is well below that
found in the aforementioned simulations. While Saturn may have formed
during the late stages of protostellar evolution when the disk mass
was severely reduced \citep{Guillot2005}, such a scenario cannot
account for the relatively low mass of several close-in and resonant
planets since it is generally accepted that their kinematic
configuration requires extensive migration through active protostellar
disks\citep{lin1996}.  In this paper, we consider another effect which
may limit the asymptotic mass of protoplanets involving the tidal
barrier. Due to its multi-dimensional nature, the complex flow pattern
is ideally analyzed with numerical simulations.

The earliest disk simulations did not have sufficient resolution to resolve
flow near the protoplanet \citep{Miki1982}. A novel approach was
introduced by \citet{Korycansky1996} to analyze inviscid flow in a
steady-state. Under these conditions, the potential vorticity
(hereafter vortensity) and the Bernoulli energy are conserved. In
their simulations, they focused their attention on a local patch close
to the protoplanet by adopting a set of boundary conditions which
approximate the nearly co-orbiting flow at large azimuthal distance
with a Keplerian velocity and smooth surface density. The advantage of
this approach is that the high resolution allows the flow pattern to
be accurately simulated \citep{Balmforth2001}. In the absence of
viscosity, however, the stream lines around the protoplanets are not
connected to that beyond the Roche radius.

Adopting the same boundary condition, \citet{Tanigawa2002} carried out
a series of high resolution time-dependent nonlinear simulations which
clearly demonstrate the presence of shock dissipation near the
accreting protoplanet and generation/dissipation of vortensity across
the shock. Nevertheless, between the shock waves, quasi steady flows
can be established with conserved vortensity ($\varpi$) and Bernoulli
energy $E_B$. They also demonstrate the possibility that a quasi
steady-state may not be attainable under some circumstances. The minor
deficiency of both these simulations was the artificial boundary
condition at large azimuth from the protoplanet. We demonstrate here
that the modification of the global disk structure can significantly
alter the accretion rate onto the embedded protoplanets.

There are many 2D global simulations of protoplanet-disk interaction
\citep[see the recent review by]{papaloizou2006}. These simulations
show that tidal torque from a sufficiently massive embedded
protoplanet not only induces dissipative shock in the disk but also
leads to gap formation
\citep[cf.]{lin1993,artymowitz1994,bryden1999,Kley1999}. There are
also 3D simulations that show gas flow into the Roche lobe of the
protoplanet \citep{bate2003,klahr2006}, and some of these simulations
have sufficient resolution to resolve the flow near the protoplanet
\citep{dangelo2002}. The accretion rate determined from these
simulations suggests that despite a significant reduction in $\Sigma$
near the orbit of the protoplanet, the growth timescale for
Jupiter-mass protoplanets is shorter than the global-disk gas
depletion timescale ($\tau_{\rm dep} \sim 3-10$ Myr) inferred from the
observation of protoplanetary disks
\citep{Haisch2001,Hartmann2005,Silverstone2006}. If the gap region is
continually replenished by diffusion into the gap, the asymptotic mass
of the protoplanets would well exceed the observed upper limit of the
planetary mass function \citep{marcy2005}.

Here, we explore the possibility that even if the gap is replenished
with tenuous gas, only a small fraction of that gas may be accreted
onto protoplanets with Roche radius ($R_R$) greater than the disk
thickness $H$. The model parameters in many previous simulations are
chosen to approximate those inferred from the minimum mass nebula
model in which the ratio between the disk thickness and radius ($H/a$)
is assumed to be around $0.07$. In these simulations, $H$ determines
the magnitude of the sound-speed and therefore the pressure gradient
in the radial direction in terms of the surface density distribution
$\Sigma$. With this assumption, $R_B$ becomes comparable to $R_R$ for
a protoplanet with $M_p \sim M_J$.

One exception to this set of parameters was the consideration of a
perturbation by a one-third Neptune-mass planet in very thin/cold disk
with $H/a \sim 0.02-0.03$ \citep{bryden2000}. In this case, $R_B$ is
several times $R_R$ and the surface density of the disk diminishes
well below that of an unperturbed disk. Although gas was not allowed
to accreate onto the core of the protoplanet, the accretion timescales
inferred from these simulations is substantially longer than those
inferred from the idealized Bondi analysis, the viscous diffusion of
the disk, and the $\tau_{\rm dep}$ inferred from the observed SED
evolution of disks in young clusters.

The disk thickness ($H$) is determined by heating from both the
intrinsic viscous dissipation and surface irradiation from the central
star. The commonly assumed aspect ratio ($H/a \sim 0.07$) at the
present-day distance of Jupiter (5.2AU) is determined under the
assumption that the disk is heated to an equilibrium temperature by
the proto-sun with the present-day luminosity of the Sun. Self
consistent models indicate that the $H/a$ at the snow line can be
significantly smaller than that assumed in the solar nebula model
\citep{garaud2006}. Taking this possibility into account, we suggest
that, in the limit of $R_R > H$, a tidal barrier imposed by a
protoplanet limits the gas flow from most regions inside the gap into
the Roche lobe. Much of the mass that diffuses into the gap
accumulates into the horseshoe orbits in the co-orbital region, but is
not able to accrete onto the planet.

In order to examine the dominant physical effects which regulate
protoplanets' asymptotic masses, we calculate the accretion rates and
flow dynamics through two sets of simulations.  Since we are primarily
interested in the limit $R_R > H$, the flow pattern is well
approximated by an 2D simulation \citep{dangelo2002}. In \S2, we
present an analysis to highlight the nature and importance of
the tidal barrier. Due to the complexity of the 2D flow, we cannot
obtain a full set of analytic solutions.  For illustration purposes,
we also adopt a grossly simplified 1D approximation for the
protoplanet's tidal potential and obtain solutions to highlight the
effect of the tidal barrier in quenching the accretion flow onto it.

In order to illustrate the structure of the flow, we present in \S3, a
set of high-resolution 2D hydrodynamic simulations to demonstrate that
the flow into the Roche lobe is indeed quenched for protoplanets with
$R_R>H$ despite diffusion of gas into the gap region. Although the 2D
simulation provides a more realistic approximation for the
multi-dimensional geometry of the potential field and the
non-spherically symmetric flow pattern of the background gas,
numerical simulations of this kind can only be carried out with a
limited resolution for several dynamical timescales. However, the disk
environment, as well as the mass growth and envelope structure of the
protoplanet, all evolve on much longer timescales, especially during
the epoch when the envelope undergoes a transition from quasi
hydrostatic contraction to dynamical accretion.  The computation of
the late stages of giant planet formation also requires following the
governing equations over large dynamical range in space and
time. While the envelope around the core adjusts to quasi hydrostatic
equilibrium, the flow around the planet's Hills radius adjusts to a
steady state. The technical challenges for 2D simulations covering
more than four orders of magnitude in physical scales over such large
timescales makes them impractical for this purpose.  In
\S\ref{sec:1dnumeric}, we show that the dominant physical effects
generated by the 2D simulations can be quantitatively captured by a
simple-to-use 1D scheme, with which we verify that the gas accretion
is significantly suppressed in the limit $R_R > H$. Utilizing this
model, we show that the flow of gas into the protoplanets' Roche lobe
declines and the protoplanets acquire asymptotic masses on Myr
timescale. Finally, we summarize our results and discuss their
implications in \S4.

\section{A Tidal Barrier for Accretion Flow}
\label{sec:section2}
In this section we begin our analysis with an analytic approximation
of the accretion flow pattern in two-dimensions. Since protoplanets
accreate from a disk environment, it is informative to compare the
disk structure with the gravitational potential of the
protoplanet. The scale height in the direction normal to the plane of
the disk is determined by the gas sound-speed $c_s$ such that
\begin{equation}
H=c_s/\Omega, 
\end{equation}
where $\Omega = (GM_\ast/a^3)^{1/2}$ is the Keplerian frequency of the
disk and $M_\ast$ is the mass of the central star. The Bondi radius
for a steady, spherically symmetric accretion flow onto a protoplanet
with a mass $M_p$ is
\begin{equation}
R_B = G M_p /2 c_s^2.
\end{equation}
The Roche radius for a planet with semi major axis $a$ is
\begin{equation}
R_R = (M_p/3 M_\ast)^{1/3} a.
\end{equation}
The magnitude of
\begin{equation}
{R_R \over R_B} = {2 H^2 \over 3 R_R^2} = 2 h_1^2
\left(1 \over 3q_3^2\right)^{1/3}
\label{eq:rrorb}
\end{equation}
and,
\begin{equation}
{H \over R_B} = {2 h_1^3\over q_3}
\label{eq:horb}
\end{equation}
where $q_3 \equiv M_p/ 10^{-3} M_\ast$ and $h_1 = c_s/(0.1 \Omega a)$
are the normalized planetary mass and disk scale height. In a minimum
mass nebula model \citep{hayashi1985} where the gas temperature is
assumed to be identical to the equilibrium temperature of the large
grains, $h_1 \simeq 0.7$ at the present radii of Jupiter ($a=5.2$ AU).

Under these conditions, the critical stage where the accretion rate is
being suppressed occurs when $R_R \sim R_B \sim H$ and $M_p = M_J$.
In principle, this region should be analyzed with full 3D
consideration \citep[cf.]{bate2003,klahr2006}. However, the inadequacy
of present-day facilities and the enormous computational requirements
of full 3D numerical simulations place severe limits on both numerical
resolution and temporal range. It is therefore useful to first
consider a more idealized 2D flow pattern.  Through numerical
simulations and analysis in this section, we emphasize that the
protoplanets' accretion rate is sensitive to the global evolution of
the disk structure. This approximation is also reasonable in the limit
that $R_R > H$ or equivalently $R_B > R_R $.

There have been multiple simulations studying the process of accretion
onto a protoplanet covering a wide range of approaches. In an attempt
to achieve high resolution near the protoplanet, several investigators
\citep{Korycansky1996,Tanigawa2002} considered flows in a local region
surrounding the protoplanet. Both studies assume that gas entering the
simulation from both the leading and trailing azimuthal boundaries has
a constant surface density and Keplerian velocity. A second set of
simulations have investigated the global disk response to
protoplanet-disk interaction.  Some of these simulations include the
protoplanet's orbital migration while others include the effect of
intrinsic turbulence. In all cases, they show that Jupiter mass
protoplanets induce gap formation in protostellar disks with
properties similar to that of the minimum mass nebula. In order to
achieve sufficient resolution for studying the flow around
protoplanets, most of these simulations are carried out for $M_J$-mass
protoplanets in disks with $H/a \sim 0.07$ such that their $R_R \sim
R_B \sim H$. Here, we are primarily interested in the disk regions
where $R_B > > R_R$ which is possible only for $R_R > H$. We show that
gas-giants with $M_p < M_J/3$ (such as Saturn) are probably formed in
regions of the disk that satisfy $H/a < 0.04$ during the epoch of
planet formation.

\subsection{Streamlines Intersecting The Feeding Zone} 
\label{section:streamlines}
The continuity and momentum equations for a two-dimensional inviscid
disk can be obtained through vertical integration such that
\begin{equation}
{\partial \Sigma \over \partial t} + \nabla \cdot \left( \Sigma {\bf
v}\right) =0
\label{eq:2dconti}
\end{equation}
and,
\begin{equation}
{\partial {\bf v} \over \partial t} + {\bf v \cdot \nabla} {\bf v} =
-2 \Omega {\bf {\hat z} {\rm x} v} - \nabla \Phi_g - c_s^2 \nabla {\rm
ln} \Sigma.
\label{eq:2deofm}
\end{equation}
Similar to the 1D analysis presented in \S\ref{sec:1dapprox}, we have
adopted an isothermal equation of state. We determine the magnitude of
${\bf v}$ in a frame which rotates at the same angular frequency
$\Omega$ as the planet orbits its host star. In the analytic
derivations, we neglect the effect of viscous and shock dissipation,
but some degree of dissipation is present in our numerical
simulations. These approximations are adequate for linking streamlines
between shock waves. We also adopt a Hill's approximation for the
potential in this frame such that
\begin{equation}
\Phi_g = -{3 \over 2} \Omega^2 x^2 - {G M_p \over (x^2 + y^2) ^{1/2} }
\label{eq:phiofg}
\end{equation}
where the shearing-sheet coordinates \citep{Goldreich1965} are
centered around the planet with $x$ and $y$ correspond to the radial
and azimuthal directions. Although this expression is a much more
accurate description of $\Phi_g$ than the one-dimensional
approximation given in Equation (\ref{eq:1dpot}), approximations must
be made to solve it analytically.

In the absence of any viscous or nonlinear dissipation, it is helpful
to analyze the flow with a vortical stream function ($\psi$) and a
scalar potential ($\phi$) which varies along the lines of constant
$\psi$ such that the 2D mass flux can be decomposed into
\begin{equation}
\Sigma {\bf v} = {\bf \nabla {\rm x} {\hat z} } \psi + {\bf \nabla}
\phi.
\end{equation}
Contours lines of constant $\psi$ and $\phi$ are orthogonal to each
other, and a steady-state ($\nabla\cdot\left(\Sigma {\bf v}\right) =
0$) corresponds to $\nabla^2 \phi =0$. The velocity is parallel to
streamlines of constant $\psi$, along which the mass flux ($\Sigma v$)
is conserved, and the momentum equation can be reduced to two
integrals of motion
\begin{equation}
{\omega + 2 \Omega \over \Sigma} = \varpi(\psi)
\end{equation}
\begin{equation}
v^2/2 + c_s^2 {\rm ln} \Sigma + \Phi_g = E_B(\psi)
\label{eq:ejacobi}
\end{equation}
which correspond to the vortensity \citep{Papaloizou1989} and
Bernoulli constant \citep{Korycansky1996,Balmforth2001}. 

Differentiating the mass flux ${\bf F}_m \equiv \Sigma {\bf v}$
and $E_B$ with respect to $\phi$ along these streamlines,
\begin{equation}
( d {\rm ln} \Sigma /d \phi )_\psi = - ( d {\rm ln} v_\psi /d
\phi)_\psi ,
\end{equation}
and,
\begin{eqnarray}
{(v_\psi^2 - c_s^2) \over \Sigma} \left( {d \Sigma \over d \phi}
\right)_\psi & = & -\left( {\partial x \over \partial \phi}
\right)_\psi {\partial \Phi_g \over \partial x} -\left( {\partial y
\over \partial \phi} \right)_\psi {\partial \Phi_g \over \partial y}\\
& = & \left( {x \over \Sigma v_x} \right) \left( 3 \Omega^2 - {G M_p
\over (x^2 + y^2)^{3/2}} \right) - \left( {y \over \Sigma v_y} \right)
{G M_p \over (x^2 + y^2)^{3/2}}
\label{eq:2dbondi}
\end{eqnarray}
where we define the magnitude of ${\bf v}$ along the streamline to be
$v_\psi$. Equation (\ref{eq:2dbondi}) clearly indicate a potential 
transonic point along each streamline.

\subsection{Analogous Bondi Solution}
\label{sec:1dapprox}
The sonic transition in a 2D flow is similar to that
in the 1D flow, and the quantities $F_m$ and $E_B$ are
equivalent to the mass flux 
\begin{equation}
\dot M_p = 4 \pi \rho u r^2
\end{equation}
and the one-dimensional Bernoulli energy
\begin{equation}
E_{B1} = u^2 /2 + c_s^2 {\rm ln} \rho + \Phi_g
\end{equation}
where $\rho$ and $u$ are the spherically symmetric density and 
radial velocity.  We find it is useful to identify the dominant physical
effects which may affect the 2D flow pattern with the help of an approximate
1D solution.

In a steady state, the equations of 1D continuity and motion are combined into
\begin{equation}
{(c_s^2 - u^2) \over u} {d u \over d r} = - {2 c_s^2 \over r} - \nabla
\Phi.
\label{eq:bondieofm}
\end{equation} 
In the standard Bondi analysis for a point-mass potential 
in which $\Phi_g = \Phi_0 \equiv - GM_p/r$, the continuity of a 
steady flow across the sonic point $r_s$ where $u=c_s$ 
requires $r_s = - 2 c_s^2 / \nabla \Phi$ and $E_B = c_s^2 {\rm ln}
\rho_\infty$. By scaling the background quantities with those at 
the sonic point $r_s$ such that $\rho_s =\rho_\infty {\rm exp} 
{3/2} \sim 4.5 \rho_\infty$, we find the accretion rate to be
\begin{equation}
\dot M_p = \dot M_{pm} \simeq 14 \rho_\infty G^2 M_p^2 / c_s^3.
\label{eq:1dmdot0}
\end{equation} 
In this case, the potential at $r_s$
is lower (more negative) than at $r=\infty$, and the gravitational
focusing effect causes $\rho_s > \rho_\infty$.

However, around a protoplanet with a circular orbit
and semi major axis $a$, the region outside its Roche radius ($R_R =
(M_p/3 M_\ast)^{1/3} a$) is dominated by the gravity of its host star.
In the frame which co-rotates with its orbit, the full tidal potential
due to both the planet and its host star is given by Equation 
(\ref{eq:phiofg}). For the 1D analysis in this section, we introduce an
approximation to the gravitational potential which includes the 
tidal effect induced by central star but neglects
its gravitational torque on the otherwise assumed spherically
symmetric (relative to the protoplanet) radial flow such that
\begin{equation} 
\Phi_g = \Phi_0 \left( 1 + \left( {R_R ^2 \over r^2}
+ {R_R ^2 \over r R_{\rm \infty} } \right)^{-1} \right).
\label{eq:1dpot}
\end{equation}
Here we did not include the Coriolis force associated with the
rotation of the frame.  In this approximation, gravity reduces to that
of a point mass ($\nabla \Phi \simeq GM_p/r^2$) for $r< R_R$, it
becomes repulsive, ($\nabla \Phi \simeq -GM_p/[R_R^2 {(1 +
r/R_\infty)}^2] + GM_p/r^2$) for $R_R < r < R_{\infty}$, and vanishes
asymptotically ($\nabla \Phi \simeq -GM_pR_\infty^2/R_R^2 r^2$) for $r
> > R_{\infty}$. In principle, this approximation cannot be used to
replace the full multi-dimensional simulation on the detailed
structure and dynamics of the outer most region of the envelope near
the protoplanet's Roche radius. However, by setting $R_{\infty}
=r_{\rm max} = {\sqrt 12} R_R$ \citep{Greenzweig1990}, the potential
in the frame which co-rotates with the protoplanet's orbit can be
reasonably approximated along the radial direction.

\begin{figure*}
\begin{center}
\plotone{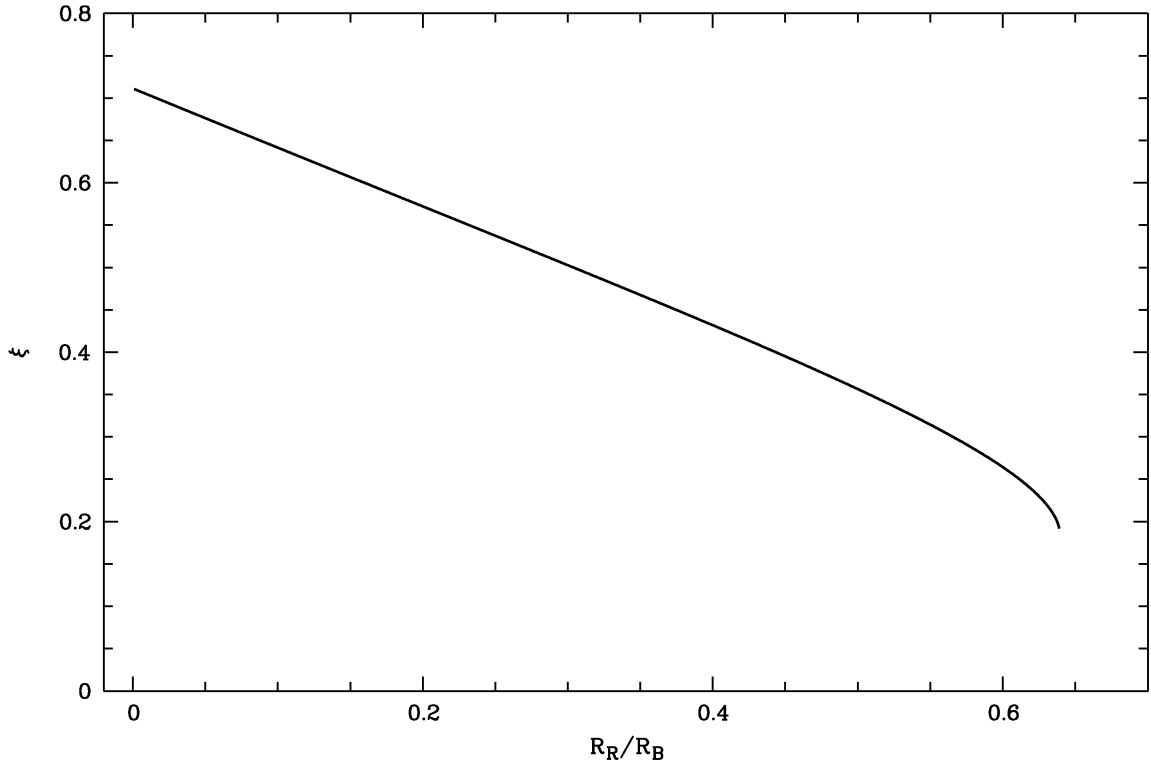}
\end{center}
\caption{Sonic radius parameter $\xi$, defined by the sonic point
$r_s=R_B/\xi$, as a function of the ratio of Roche to classical Bondi
radii. The dependence of $\xi$ on $R_R/R_B$ can be found in Equation
(\ref{eq:sonicparam}).}
\label{Fig:quadratic}
\end{figure*}

For the tidal potential, the
right hand side of Equation (\ref{eq:bondieofm}) vanishes at $r_s
\equiv R_R/ \xi$ where the parameter $\xi$ can be obtained from a
quadratic equation
\begin{equation}
\xi^2 + \left({2 R_R \over R_\infty } + {R_R \over R_B} \right) \xi +
\left( {R_R^2 \over R_\infty^2} +{2 R_R^2 \over R_\infty R_B} -1 \right)
+{R_R^3 \over R_B R_\infty^2} {1 \over \xi} =0
\label{eq:sonicparam}
\end{equation}
the solution of which is plotted in Figure (\ref{Fig:quadratic}). In the
limit that $R_R < R_B$, $\xi \simeq 1 - R_R/R_\infty + R_R/2 R_B$,
i.e. the sonic point is close to the Roche radius.
\clearpage
\clearpage

When the tidal potential is taken into account, $E_B = c_s^2 {\rm ln}
\rho_\infty - G M_p R_\infty/R_R^2$ at large $r$, such that the
density at the sonic point is given by
\begin{equation}
\rho_s = \rho_\infty{\rm exp} \left[\frac{\Phi(r_s) -
\Phi(\infty)}{c_s^2} -\frac{1}{2}\right].
\end{equation}
Since the absolute value of the potential at $r_s$ is near a maximum
(where gravity $\nabla \Phi \sim 0$), $\rho_s < \rho_\infty$. This
density decline is due to the need of a positive pressure gradient,
required to overcome the tidal barrier associated with this potential
maximum. In the limit of relatively small $c_s$ (or equivalently large
$R_B$) a large density reduction is needed. Based on the above
extrapolation that $r_s \sim R_R$ for $R_R < < R_B$,
\begin{equation}
\rho_s \simeq 0.6 \rho_\infty {\rm exp} \left(- \eta R_B/R_R\right)
\end{equation} 
where $\eta \equiv 2 (R_\infty/R_R -2) \sim 3$ is positive. In a
tidal potential with $R_R < < R_B$, $\xi\sim 1$, $r_s \sim R_R$, and
\begin{equation}
\dot M_p = \dot M_{t} \simeq \left( {2 \pi \over 3^{1/3}}
\right) c_s R_R^2 \rho_\infty 
{\rm exp}\left(-\eta{R_B\over R_R} \right) =
0.135 \dot M_{pm} \left({R_R \over R_B} \right)^2 {\rm exp} \left(-\eta{R_B
\over R_R} \right) .
\label{eq:1dmdot}
\end{equation}
The above equation indicates that, in the limit $R_R < < R_B$, the
tidal accretion rate ($\dot M_{t}$) can be much smaller than the
standard Bondi accretion rate for a point mass potential $\dot
M_{pm}$.

\subsection{Boundary Conditions in 2D Flows: Loading the Streamlines}
\label{sec:loading}
A comparison of the governing equations (\ref{eq:bondieofm}) and 
(\ref{eq:2dbondi}) indicates that the competing
effects of pressure gradient and gravity  in the 2D flows is determined
by the boundary conditions of the streamlines.  At large $x$
distance from the planet, the unperturbed 2D flow is approximately
Keplerian, with $v_x \sim 0$ and $v_y \sim -3 \Omega x/2$, which can
be substantially larger than the sound-speed. These streamlines are on
hyperbolic paths with large impact parameters relative to the
protoplanet and material on these streamlines will not accreate onto
the planet.

The streamlines which are relevant to accretion flow are those with
impact parameters smaller than $R_R$. In the limit $H < < R_R$,
pressure does not contribute significantly, and Equation
(\ref{eq:2deofm}) reduces to that for non-interacting
particles. Particles that originate from unperturbed Keplerian orbits
at $x_\infty = x_{\rm max} \equiv \sqrt{12} R_R$ and $y_\infty = \pm
\pi a$, reach the vicinity of the unstable Lagrangian saddle points
$(\pm R_R, 0)$ with negligible velocity \citep{Greenzweig1990}. The
region of the disk between $a-x_\infty$ and $a+ x_\infty$ is commonly
referred to as the feeding zone.  Although particles with initial
Keplerian orbits that pass through the location $y_\infty = \pm \pi a$
with $x_\infty = 1.8x-2.7 R_R$ can directly strike a compact
protoplanet core, particles' on neighboring orbits are strongly
scattered by the planet \citep{bryden2000}.

Although $\varpi$ and $E_B$ are useful quantities for analyzing the
flow in the horseshoe region, they are not conserved for gas accreted
from the disk onto the planet. More specifically, for fluid in the
protostellar disk with Keplerian asymptotic velocities,
$\varpi=\Omega/2 \Sigma_\infty$. However, if gas enters into the
planet's Roche radius with $x^2+y^2 < 3 R_R^2/4$ in a Keplerian
circumplanetary disk, its streamlines have $\omega < - 2 \Omega$ and
$\varpi < 0$. This mismatch in $\varpi$ implies that gas can only be
accreted from protostellar disks onto protoplanetary disks through
shock dissipation \citep[see 2D simulations by]{Tanigawa2002}. Since
any dissipation would invalidate the conservation of $\varpi$ and
$E_B$, the flow pattern derived from these conservative quantities
cannot adequately describe accretion flow onto the protoplanet
\citep[cf.]{Korycansky1996}.

However, these quantities are useful for determining the conditions
under which gas on some streamlines in the horseshoe regions may avoid
shock dissipation and accrete onto the protoplanet.  The flow in this
region depends sensitively on the boundary condition at $(x_\infty,
y_\infty)$. If the flow at $(x_\infty, y_\infty)$ is strictly
Keplerian, as assumed in some previous local simulations, $E_B = c_s^2
{\rm ln} \Sigma_\infty - 3 \Omega^2 x_\infty^2/8$. In this case,
\begin{equation}
\Sigma_s \equiv \Sigma (x_s, y_s)=\Sigma_\infty {\rm exp}\left[- \beta
{R_R^2 \over H^2} - {1 \over 2}\right]
\label{eq:2dsigma}
\end{equation}
at the sonic point $(x_s, y_s)$ where $u = c_s$. The parameter $\beta$
is independent of $H$ or $c_s$. For Keplerian boundary conditions, the
value of
\begin{equation}
\beta = {3 \over 2} \left[ \left( {x_\infty ^2 \over 4 R_R^2}
\right) - {x_s^2 \over R_R^2} - {2 R_R \over (x_s^2 + y_s)^{1/2} 
} \right] .
\label{eq:ebkepler}
\end{equation}
In this expression for $\beta$, the actual values of $(x_s, y_s)$ can
only be determined by the detailed solutions of Equations
(\ref{eq:2dconti}) and (\ref{eq:2deofm}).

In analogy with the 1D Bondi solutions, we suggest that there may
exist a sonic transition point in Equation (\ref {eq:2dbondi}) as is
found in Equation (\ref{eq:bondieofm}).  In this case, if $\vert u
\vert < c_s$ at some point along the streamline, the right hand side
of Equation (\ref{eq:2dbondi}) would vanish during the sonic
transition at $(x_s, y_s)$ with
\begin{equation}
{v_{x} (x_s, y_s) \over v_{y} (x_s, y_s)} = {x_s \over y_s}\left(
{\left(x_s^2 + y_s^2\right)^{3/2} \over R_R^3} -1 \right).
\end{equation}
The point $(x_s, y_s)$ is close to the turn-around point (where $u_y$
vanishes) of the horseshoe orbit for non-interacting particles.  Since
$v_x^2 + v_y^2 = c_s^2$ at $(x_s, y_s)$, we can express,
\begin{equation}
{v_y^2 \over c_s^2} = \left[ 1 + \left( {\left(x_s^2 +
y_s^2\right)^{3/2} \over R_R^3} -1 \right)^2 \left({x_s \over
y_s}\right)^2 \right]^{-1}.
\label{eq:vy2}
\end{equation}

In the 1D approximations (see the previous subsection), the physical
parameters at the sonic points can be analytically determined from
those at infinity.  But in 2D flows, Equations (\ref {eq:2dbondi}) and
(\ref{eq:vy2}) generally do not lead to an analytic relation between
$r_s$ and $c_s$. However, their approximate solutions suggest that the
condition for $v_x^2< c_s^2$ and $v_y^2< c_s^2$ is $r_s^2 \equiv x_s^2
+ y_s^2 \sim R_R^2$, {\it i.e.}  in the 2D limit, the transonic radius
is also comparable the planet's Roche radius ($r_s \sim R_R$) and
\begin{equation}
\dot M_p \simeq 2 \pi R_R c_s \Sigma_s = 2 \pi
R_R H \Omega \Sigma_\infty {\rm exp} \left[ - \beta \left({R_R^2
\over H^2} \right)- {1 \over 2} \right].
\label{eq:2dmdot}
\end{equation}
The above expression clearly indicates the importance of $\beta$. If
$\beta<0$, $\dot M$ will decrease with $H$ as shown in the local
simulations of \citet{Tanigawa2002}. However, if $\beta >0$, $\dot M$
will increase with $H$ as shown in the global simulation by
\citet{bryden1999}.

We suggest this dichotomy can be attributed to the boundary conditions
at ($x_\infty, y_\infty$). In order to illustrate this conjecture, we
first consider a special streamline which passes through a sonic
transition at the same azimuthal phase as the planet ({\it i.e.}
$y=0$), the regularity condition for Equation (\ref{eq:2dbondi})
requires $\partial \Phi_g/\partial x = \partial \Phi_g/\partial y=0$
or equivalently, $x_s = R_R$. In the limit that $H < < R_R$, $\Phi_g
(x_s, y_s)=-9\Omega^2 R_R^2/2$, $x_\infty = x_{\rm max} = {\sqrt 12}
R_R$, $y_\infty = \pi a$, and $\beta \sim 0$. These streamlines are
analogous to the parabolic trajectories of those particles which can
marginally reach the separatrix at the Lagrangian points from an
initial circular Keplerian orbit. Starting with Keplerian motion,
streamlines with $x_\infty > x_{\rm max}$ do not enter into the
planet's Roche lobe. However, Keplerian streamlines with $R_R <
x_\infty < x_{\rm max}$ can enter into the planet's $R_R$ and cross
the sonic radius with very small negative values of $\beta$. In local
simulations done previously, the disk flow at some specified
boundaries is explicitly set to be Keplerian at an azimuthal phase
$y_\infty < \pi a$. This artificial choice of boundary condition leads
a small negative value of $\beta$ for disks with both negligible and
finite $H$'s. This result implies that tidal barrier is overcome
purely by the initial kinetic energy rather than the positive pressure
gradient and $\Sigma_s$ is slightly larger than $\Sigma_\infty$ as a
consequence of convergent flow. If $\beta$ is negative, Equation
(\ref{eq:2dmdot}) indicates that $\dot M_p$ is a decreasing function
of $H$. This is in agreement with the simulations of
\citet{Tanigawa2002}. With their value of $\Sigma_\infty$ ($\sim
\Sigma_{mn}$), the growth timescale would be $\tau_p = M_p / \dot M_p
\sim 10^{3}{\rm yr}(M_p/10 M_\oplus)^{-0.3}$, which is much shorter
then the global depletion timescale ($\tau_{dep}$) for the disk. The
discrepancy between these timescales illustrates the need for a
mechanism to quench the accretion flow.

In contrast to local simulations, full-scale 2D simulations the
protoplanet induces the formation of a gap in the disk through both
accretion onto itself and tidal dissipation within the disk
\citep{bryden1999}. In such simulations the boundary conditions at
($x_\infty, y_\infty$) are significantly different from those assumed
in local simulations. Near the edge of the gap, the perturbed pressure
gradients cause the velocity of the gas to become super Keplerian at
positive $x_\infty$ and sub Keplerian at negative $x_\infty$. The
magnitude of the flow velocity,
\begin{equation}
v_y (x_\infty, y_\infty) = \Omega a\left( {H^2 \over 2a^2} {\partial
{\rm ln} \Sigma \over \partial {\rm ln} x} - {3 x \over 2a}
\right)_{x=x_\infty},
\label{eq:vyinfty}
\end{equation}
is reduced for both positive and negative value of $x_\infty$. The
typical gap has a width $R_R < \Delta < x_{\rm max}$. When the thermal
and viscous conditions for gap formation ($R_R > H$ and $M_p/M_\ast >
40 \nu/\Omega a^2$ where $\nu$ is the effective viscosity) are
satisfied, the surface density inside the gap $\Sigma_\infty
(x_\infty, y_\infty) \sim \epsilon \Sigma_o$ with $\epsilon <
10^{-2}-10^{-4}$ \citep{bryden1999}. Although this reduction in
$\Sigma_\infty$ decreases the value of $\dot M_p$ (by a factor similar
to $\epsilon$), the growth timescale $\tau_p$ would still be more than
an order of magnitude shorter than $\tau_{\rm dep}$ if the protoplanet
is able to accrete all the gas which diffuses into the gap region
either through turbulence \citep{Nelson2003} or due to intrinsic or
artificial viscosity \citep{artymowitz1994,Kley2001}.

In the limit of severe surface density depletion near the protoplanet,
the magnitude of $v_y$ within the gap may be substantially reduced
from the Keplerian approximation. In the limit $\vert x_\infty \vert <
H$, we approximate Equation (\ref{eq:vyinfty}) as $v_y^2 (x_\infty,
y_\infty) = (9/4) \Omega^2 x_\infty^2 (1 - f_{\rm sub})$ where
\begin{equation}
f_{\rm sub} = {2 \over 3} {H \over x} {H \over a} {\partial {\rm ln} \Sigma
\over \partial {\rm ln} x}.
\label{eq:fsub}
\end{equation}
For values of $0 < f < 1$, the Bernoulli constant can be written as
$E_B \simeq c_s^2 {\rm ln} \Sigma_\infty - 3 (1 + 3f_{\rm sub})
\Omega^2 x_\infty^2/8$. At large azimuthal phases, the streamlines
(constant $\psi$) are nearly parallel to the $y$-axis. Along these
streamlines, the potential has a maximum at $\partial \Phi_g/\partial
x = 0$, which is located on a circle $x^2 + y^2 = R_R^2$ centered on
the protoplanet. In order for the background gas to reach the
protoplanet within the Roche radius, it must pass through this local
potential maximum with a positive pressure gradient. In the limit of
low sound-speed, a sufficiently strong pressure gradient is only
attainable with a large density gradient, {\it i.e.}  the severe gas
depletion interior to $R_R$. For these conditions, the value of
\begin{equation}
\beta = 3\left[ {1 \over 8} \left( {(1 + 3 f_{\rm sub}) x_\infty ^2 
- 4 x_s^2 \over R_R^2} \right) - 1 \right].
\label{eq:beta2}
\end{equation}
In the limit $x_\infty \simeq x_{\rm max}$ and $x_s \sim R_R$, $\beta
\simeq 27 f_{\rm sub} /2 > 0$. For any value of $0 < f_{\rm sub} < 1$,
$\beta > 0$ for all streamlines with $x_\infty > \left(\sqrt 12/(1+3
f_{\rm sub})\right) R_R$. Streamlines with smaller values of
$x_\infty$ attain horseshoe orbits that do not enter into the Roche
radius. In cold disks ($H<R_R$) this tidal barrier suppresses the flow
in the azimuthal direction such that only a small fraction of the gas
which is diffused into the gap can actually reach its Roche
lobe. Consequently, the accretion rate derived from Equation
(\ref{eq:2dmdot}) become substantially smaller than that obtained when
neglecting the effect of the azimuthal tidal barrier.

\subsection{Protoplanet's Asymptotic Mass}

We now apply the appropriate boundary condition in protostellar disks
to determine the asymptotic mass of protoplanets. For illustration
simplicity, we first adopt the 1D approximation and then consider the
full 2D flow.

The density of the ambient gas can be derived from $\rho_\infty =
\Sigma / 2 H$ where the surface density is $\Sigma= f_g \Sigma_{mn}$
and $f_g$ is a scaling factor. We use the minimum mass nebula model
$\Sigma_{mn} \sim \Sigma_0(a/1{\rm AU})^{-3/2}$, where
$\Sigma_0\left(t_0\right)=2 \times 10^3 g/cm^3$ as a fiducial
prescription. We also approximate the evolution of the disk by
$\Sigma= \Sigma_0 {\rm exp} (- t / \tau_{\rm dep})$ \citep{ida2004_1}
where the characteristic gas depletion timescale $\tau_{\rm dep}$
inferred from the observed SED is $\tau_{\rm dep}\sim 3-10 Myr$.

From the 1D mass accretion rate given in Equation (\ref{eq:1dmdot}),
we can determine the asymptotic mass of the planets $M_f$ by first
solving for $x_f = 3^{1/6} (\eta/2)^{1/2} h_1^{-1} q_{3f} ^{1/3}$ with
\begin{equation}
\int_0 ^{x_f} {\rm exp} x^2 dx = A_0 {f_g \Sigma_{mn} a^2 \over h_1
M_\ast} {\tau_{\rm dep} \over P}
\label{eq:mfinal0}
\end{equation}
where $A_0 = \left(10^8 2^3/3\right)^{1/2} \pi^2 \eta^{1/2}$ is a
constant, and the planet's asymptotic mass is contained in $q_{3f}
\equiv M_f/10^{-3} M_\ast$. The left hand side of Equation
(\ref{eq:mfinal0}) is a rapid increasing function of $x_f$. For a
minimum mass nebula model ({\it i.e.}  $f_g =1$ at a=5 AU) the
magnitude of its right-hand side is $\sim 10^6$ and the solution of
Equation (\ref{eq:mfinal0}) is $x_f \sim 4$. However, the protoplanet
may be embedded inside a gap where the gas density is reduced by at
least 2 orders of magnitude from that in the minimum mass nebula. Long
period protoplanetary embryos may also have formed during the advanced
stages of disk evolution when the magnitude of $f_g$ may also be
relatively small. However, the value for $x_f$ is relatively
insensitive to the magnitude of $f_g$ and it reduces to $\sim 3$ if
$f_g \sim 10^{-2}$. The corresponding value for the asymptotic planet
mass is
\begin{equation}
M_f \simeq {x_f ^3 \over 3^{1/2} } \left( {2 \over \eta} \right)^{3/2} 
\left( { H \over a} \right)^3 M_\ast.
\label{eq:mfinal1}
\end{equation}

The structural parameters of the disk are functions of $M_\ast$ and
$a$. Near the snow line of a minimum mass nebula, the magnitude of
$M_f$ is comparable to that of Jupiter. Equation (\ref{eq:mfinal1}) is
also equivalent to $R_R (M_f) \sim H$. This relation is in agreement
with the thermal condition for the tidal truncation and gap formation
of protostellar disks by embedded protoplanets \citep{lin1986}.

The dependence of $\Sigma_s/\Sigma_\infty$ on the sign of $\beta$ in
2D is similar to that of $\rho_s/\rho_\infty$ on the presence of the
tidal barrier in the 1D approximation. Using the 2D mass accretion
rate given by Equation (\ref{eq:2dmdot}), together with the
approximate global $\Sigma$ depletion formula, we deduce an asymptotic
mass for a protoplanet of
\begin{equation}
{M_p \over M_\ast} = {3 \over \vert \beta \vert^{3/2}} \left( {H \over
a} \right)^3 \left[ {\rm ln} {8 \pi \vert \beta \vert \over 9} \left(
{a M_{\rm gap} \tau_{\rm dep} \over H M_\ast P} \right) \right]^{3/2}
\end{equation}
where $M_{\rm gap} = \pi\epsilon\Sigma_o a^2$ is the characteristic
mass associate with the gap region. For $\beta \sim 1$ and $\epsilon
\sim 10^{-3}$, the above asymptotic mass would yield $R_R \sim
H$. However, if the magnitude of $\beta$ is very small or negative,
the above equation reduces to $M_p / M_\ast \sim 3 (8 \pi H M_{\rm
gap} \tau_{\rm dep} / 9 \pi a M_\ast P)^{3/2}$, which is substantially
larger. To calculate the actual values of $\beta$ and $f_{sub}$, we
must utilize numerical simulations

\section{Two Dimensional Numerical Simulations}

\subsection{Numerical Method}
In order to verify the 2D analytic approximation in the previous
section and deduce a value for $\beta$, we now present a series of 2D
numerical simulations. The 2D numerical scheme we use is a fully
parallel hydrodynamical code with which the continuity and momentum
equations are solved on a fixed Eulerian grid in 2D cylindrical
coordinates $\left(r,\phi\right)$. A staggered grid is introduced
where scalars are defined at grid centers and vectors are defined on
grid edges, allowing us to maintain second-order spatial
accuracy. Earlier versions of the code were used, among other
applications, to study the formation of gaps in protostellar disks
\citep{bryden2000} and also boundary layer physics in the accretion
disks around cataclysmic binaries \citep{kley1989}. The equations of
continuity and motion for the fluid and the expression for the
gravitational potential are similar to those given in Equations
(\ref{eq:2dconti}), (\ref{eq:2deofm}), and (\ref{eq:phiofg})
respectively. In the equation of motion, we adopt an isothermal
equation of state $P=c_s^2\rho$, with a radially varying sound-speed,
$c_s = (H/a) V_{kep}$ where the aspect ratio of the disk $H/a$ is
assumed to remain constant. In order to make a direct comparison with
the results of the analytic approximation, we neglect the effect of
the viscous stress associated with the turbulent motion. A softening
parameter $r_{soft}=0.5R_R$ is adopted for the gravitational potential
of the planet.

The planet is initialized with a $1 M_{jup}$ planet at a location
$\left(r,\phi\right)=\left(1AU,0\right)$. Although the planet is held
fixed, any mass accreted during the simulation adds to the overall
gravitational potential of the planet. The simulations cover the
entire azimuthal extent of the disk and a radial range from $0.5AU$ to
$1.5 AU$. Simulations were initialized with a surface density profile
of $\Sigma=950 \frac{g}{cm^3}\left(\frac{r}{1 AU}\right)^{-3/2}$, and
have a numerical resolution of $\left(r,\phi\right) =
\left(200,1600\right)$.

Each of the terms in the fluid equations are dealt with consecutively
through both operator and directional splitting. Our advection scheme
is an extension of the simple first-order upwind scheme based on the
prescription of \citet{vanLeer1977}, and allows for the detailed
treatment of strong shocks. Details of the advection scheme can be
found in \citet{kley1987}. We solve the entire problem in a frame
rotating with the planet.

\subsection{Two Dimensional Numerical Models}
Before the fluid is incorporated onto the planet it must navigate
though both the shocks and sub-disk surrounding the protoplanet. In
the 2D study we neglect any radiative feedback onto the gaseous disk
from the energy released during the cooling and contraction of the
planetary atmosphere. Following \citet{Tanigawa2002}, we decrease the
surface density of the disk in the region $r_{sink}\le 0.5 R_R$ at a
rate given by
\begin{equation}
\Sigma\left(t+\Delta t\right) =
\Sigma\left(t\right)\left[1-\frac{\Delta t}{\tau_{sink}}\right].
\end{equation}
This mass is added to the mass of the protoplanet, and gradually
increases its gravitational contribution. \citet{Tanigawa2002}
demonstrated that the overall accretion rate ($\dot{M}$) was
relatively unaffected by the choice of $\tau_{sink}$ provided that
$r_{sink}\approx 0.1R_R$. Because of our lower numerical resolution
near the protoplanet we set $r_{sink}=0.5R_R$ and
$\tau_{sink}=1/\Omega$. The dominate effect which regulates the
accretion rate is the supply of material to the sub-disk region. This
quantity is determined by the effectiveness of loading the streamlines
far from the planet, so the values of $r_{sink}$ and $\tau_{sink}$
should not be crucial in determining the asymptotic mass of the
planets. In order to test the validity of this assumption and the
convergence of our results, we carried out several simulations with
varying values of $\tau_{sink}$, but these showed little variation in
the overall accretion rate.

\begin{figure*}
\begin{center}
\plotone{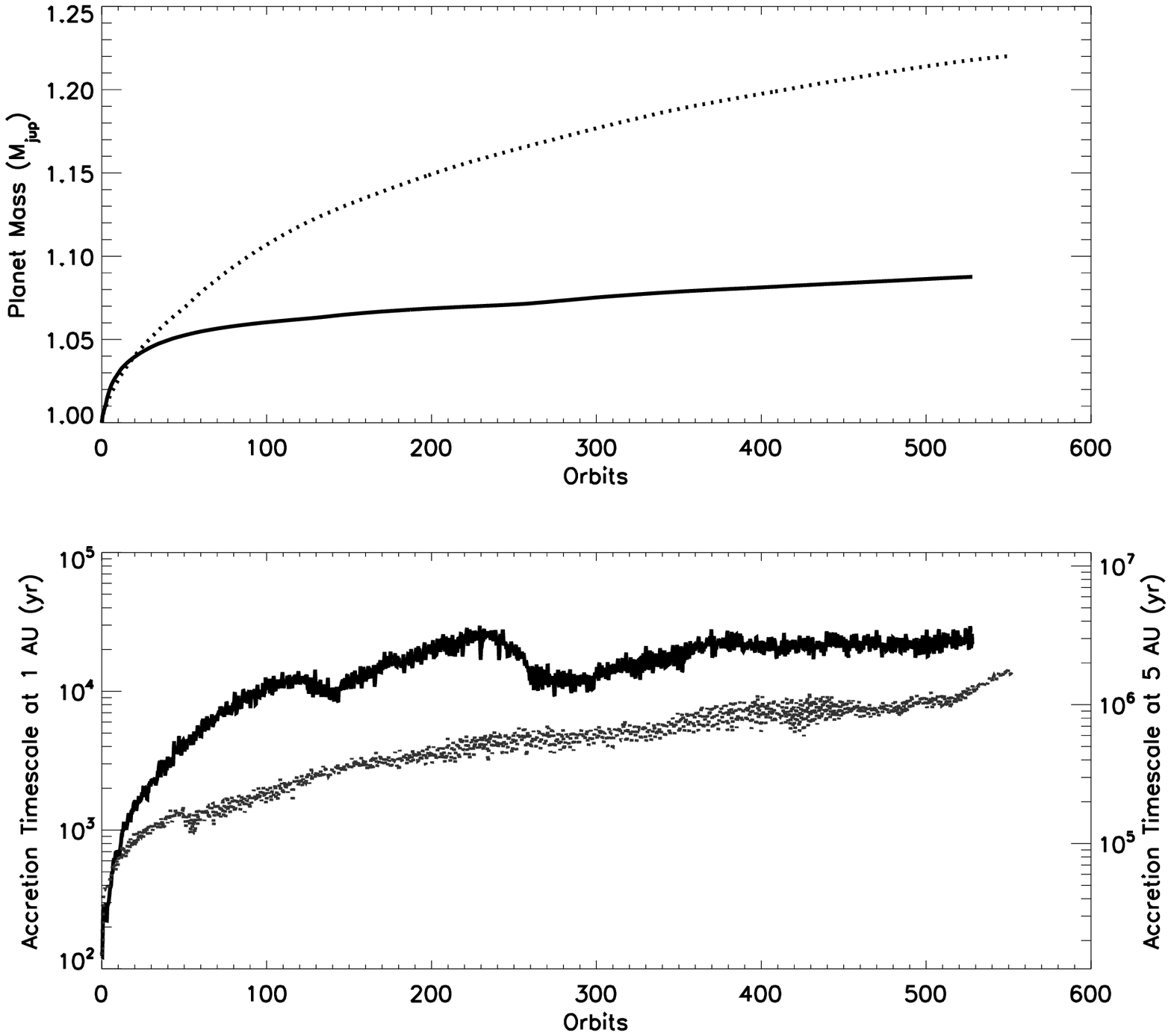}
\end{center}
\caption{The total mass of the planet as a function of time (top). The
solid line shows results for $\left(\frac{H}{a}\right) = 0.04$, while
the dotted line shows $\left(\frac{H}{a}\right) = 0.07$. The lower
panel shows the accretion timescale for both simulations. The
left-hand ordinate shows the timescale for a planet at 1 AU, while the
right-hand ordinate shows the timescale for a planet located at 5 AU.}
\label{Fig:mplanet}
\end{figure*}

We present the results of two models in which $H/a=0.04$ and $0.07$
respectively. Figure (\ref{Fig:mplanet}) shows both the total mass and
accretion timescale as a function of orbital period for these
models. The accretion rates approach a steady state on the time scale
of $100$ orbital periods.  This time scale is comparable to the
liberation period of flows in the horseshoe region and the synodic
period of the flow near the separatrix at the Lagrangian points.  It
reflects the protracted adjustment of the disk structure to the tidal
perturbation of the protoplanet. The accretion timescale ($\tau_{\rm
1AU} = M_p/ \dot M_p)_{\rm 1 AU}$ for a Jupiter-mass protoplanet is
proportional to both the local surface density ($\propto a^{3/2}$) and
to its orbital period ($\propto a^{3/2}$). It can therefore be scaled
to other radii such that
\begin{equation}
\tau_{\rm growth} = \tau_{\rm 1 AU} (a/{\rm 1 AU})^3.
\end{equation}
Figure (\ref{Fig:mplanet}) shows timescales for planet located at both
$1$ and $5AU$. For a planet at $5AU$, we see that the growth timescale
becomes comparable to the gas depletion time scale of the disk. Within
the computational duration, the total fractional gain in the planetary
mass is 0.05 and 0.2 for $H/a =0.04$ and 0.07 at 1 AU and one order of
magnitude smaller at 5 AU. Protracted 2D simulations for protoplanets'
nonlinear growth (over $10^4$ orbits) require computational resources
beyond that currently available.

\begin{figure*}
\begin{center}
\plotone{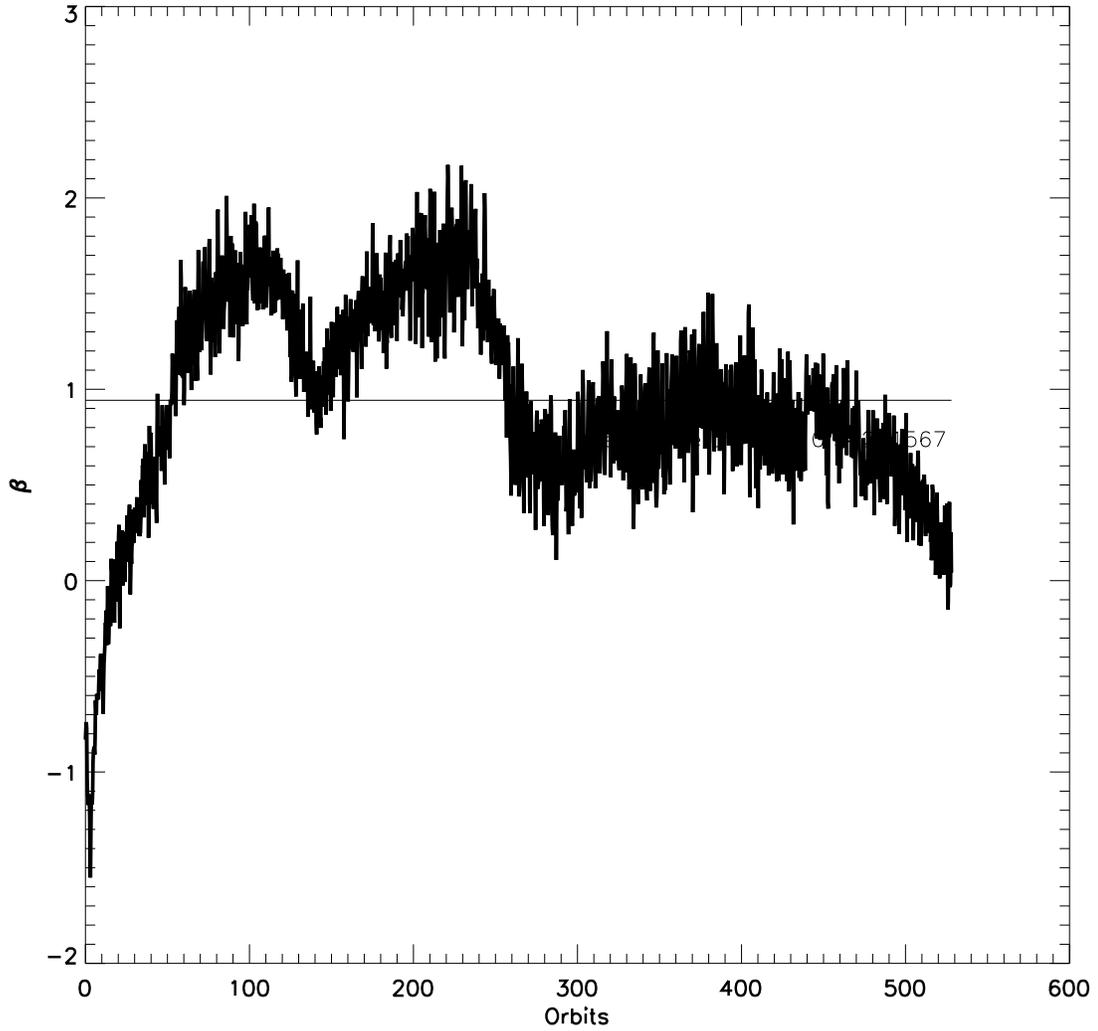}
\end{center}
\caption{The evolution of $\beta$, deduced from the comparison of two
models with $\left(\frac{H}{a}\right) = 0.04$ and 0.07. From Equation
(\ref{eq:2dmdot}), we see that a positive $\beta$-value indicates that
$\dot M$ is an increasing function of the disk scale height ($H$).}
\label{Fig:beta}
\end{figure*}

The results in Figure (\ref{Fig:mplanet}) also clearly indicate that
the mass of a planet embedded in a disk with a lower aspect ratio
levels off at significantly lower mass than one embedded in a disk
with larger $(H/a)$. Based on these results, we compute, from Equation
(\ref{eq:2dmdot}), the value of
\begin{equation}
\beta \simeq \left( {R_R^2 \over H_2^2} - {R_R^2 \over H_1^2}
\right)^{-1} {\rm ln} \left( {\dot M_1 H_2 \over \dot M_2 H_1 }
\right)
\end{equation}
where the subscripts 1, 2 refer to the values for the two models. The
value of $\beta$ increases from a very small value to the order of
unity after $\sim 50$ orbits (see Figure (\ref{Fig:beta})). The disk
is initialized as a Keplerian disk, and as a result there is some
amount of mass on Keplerian orbits within the separatrix region
$\left(R_R<x<\sqrt{12}R_R\right)$ at azimuthal phase $y_\infty<\pi
a$. This gas passes over the tidal barrier near the planet solely
based on this artificial kinetic energy. For this reason the first
$\sim 30$ orbits in Figure (\ref{Fig:mplanet}) show similar accretion
rates onto the planets in both simulations and the initial value of
$\beta$ is more closely approximated by Equation
(\ref{eq:ebkepler}). In fact, during this relaxation phase, the lower
sound-speed simulation is able to accrete mass onto the planet faster
because the outward pressure gradient that develops around the planet
is smaller.

\begin{figure*}
\begin{center}
\plotone{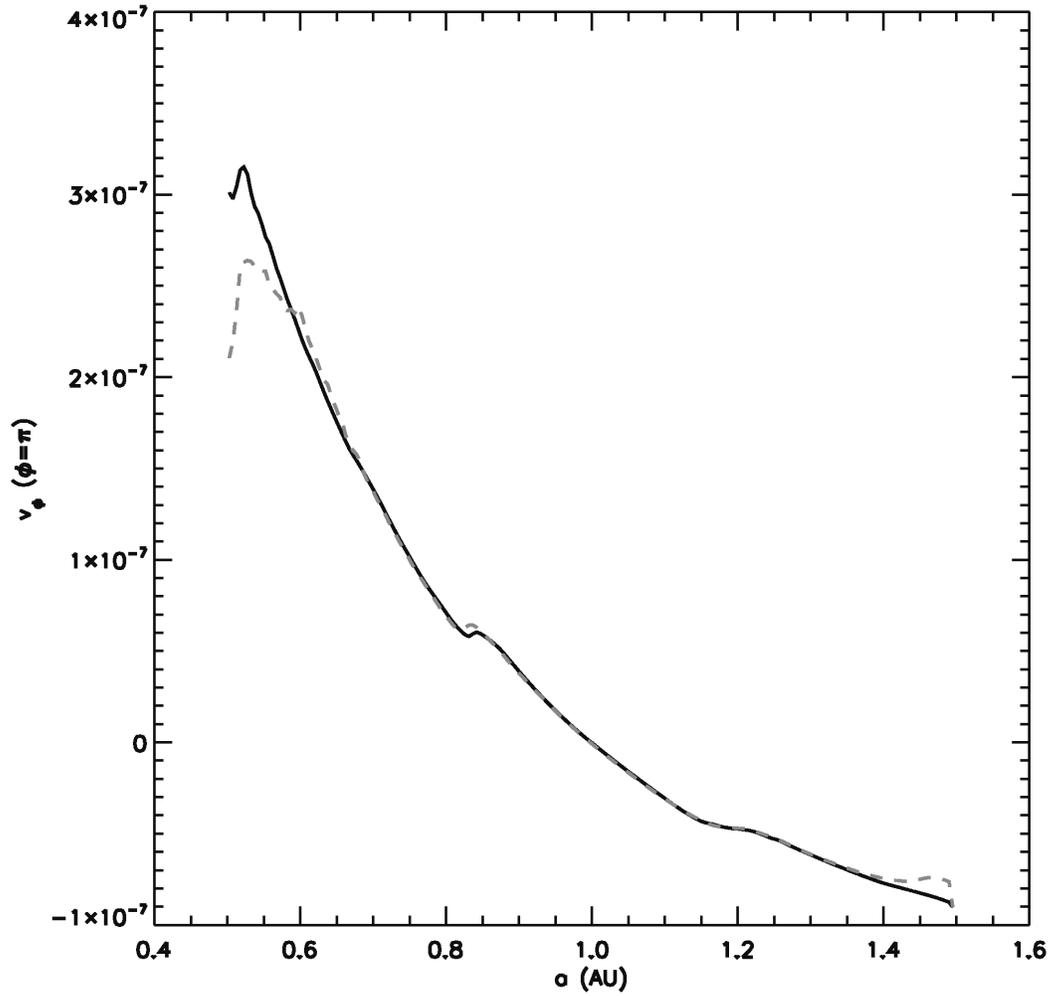}
\end{center}
\caption{The radial distribution of $v_y$ at the protoplanet's
superior conjunction for models with $\left(\frac{H}{a}\right) = 0.04$
(solid line) and $0.07$ (dashed line). A decline in the magnitude of
$v_y$ within the feeding zone is due to a pressure gradient associated
with the surface density distribution across the gap.}
\label{Fig:vygap}
\end{figure*}

After approximately one libration period around the L$_4$ and L$_5$
points, mass elements that start their trajectory at the superior
conjunction ({\it i.e.} $y= a \pi$) can reach the vicinity of the
protoplanet's Roche radius. Thereafter, a large pressure gradient
along the stream line is needed to overcome the tidal potential
barrier. During this second phase, the accretion timescale can be up
to an order of magnitude larger in the high sound-speed case. The
value of $\beta\sim 1$ is more closely approximated by Equation
(\ref{eq:beta2}). The discussion near Equation (\ref{eq:vyinfty})
indicates that this approximation is based on the assumption that $v_y
(x_\infty, y_\infty) \simeq 0$. In Figure (\ref{Fig:vygap}), we plot
the value of $v_y$ at the planet's superior conjunction as a function
of the disk radius. We note that within the feeding zone, the
magnitude of $v_y$ is significantly reduced from its Keplerian values
which is in agreement with Equation (\ref{eq:vyinfty}).
\clearpage

\begin{figure*}
\begin{center}
\plotone{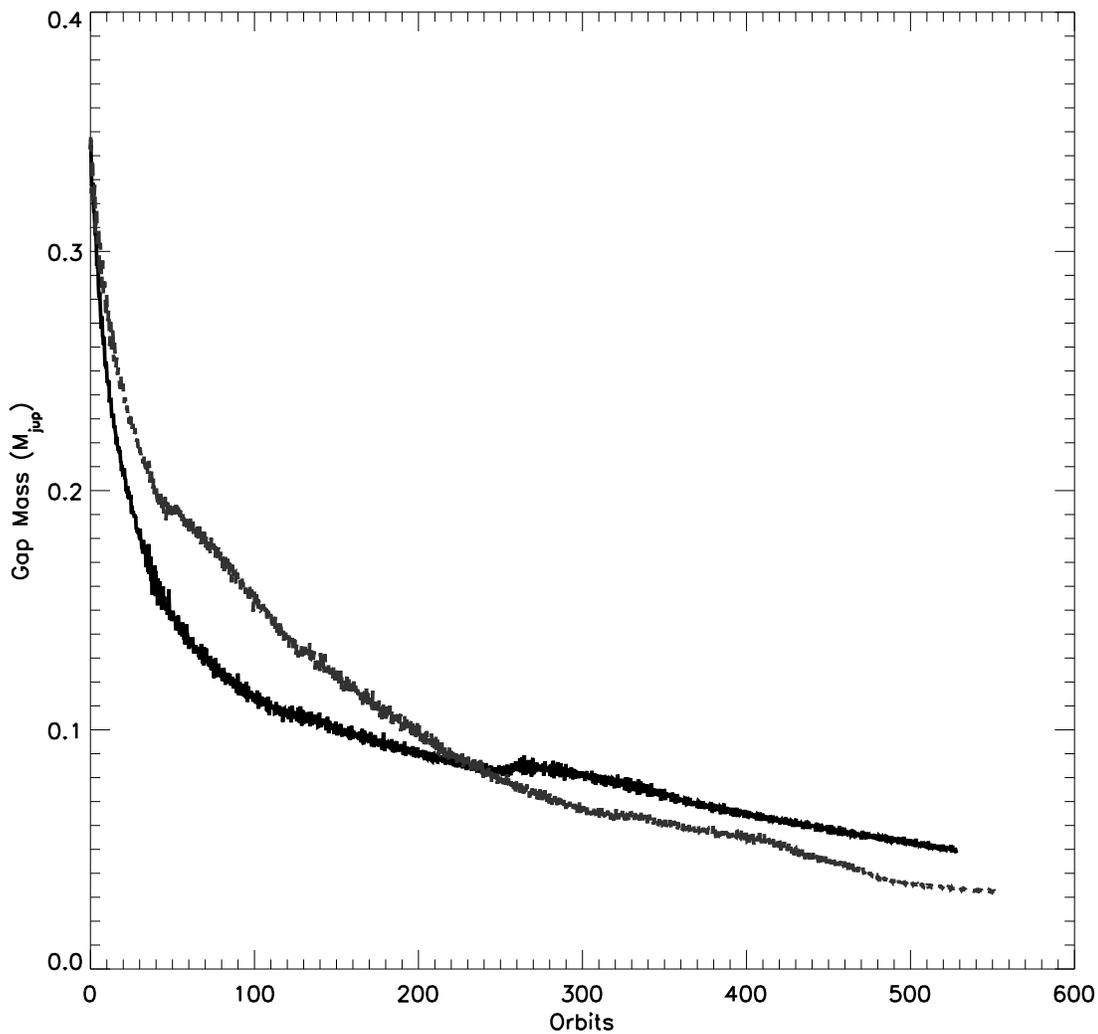}
\end{center}
\caption{The total mass of the planet in the gap region $\left(|r-a|
\le \sqrt{12}R_R\right)$ as a function of orbits for $\frac{H}{r} =
0.04$ (solid line) and $\frac{H}{r}=0.07$ (dashed line). By the
end of the simulation there is more mass within the gap for the low
sound-speed case, indicating that the tidal barrier, not the lack of
material within the feeding zone, is the cause of the planets
increasing growth timescale.}
\label{Fig:mgap}
\end{figure*}

\clearpage
Our simulations do not include explicit viscosity, and as a result
mass diffusion into the gap is due to numerical viscosity or
shocks. Figure (\ref{Fig:mgap}) shows the total mass within the
separatrix region $\left(|r-a|\le \sqrt{12}R_R\right)$ for the two
simulations. Again, during the initial stages of the simulation the
two cases track each other. The drop in gap mass is quite precipitous
($\sim 0.3M_{Jup}$) during the first $100$ orbits. Comparison with
Figure (\ref{Fig:mplanet}) shows that a majority of this gas is not
accreted by the planet, but rather shocked by the planet and pushed
out of the region. However, the lack of material is not the cause for
the mass of the protoplanet to level off in the low sound-speed model;
by the end of the simulation, there is more mass within the gap region
in this simulation than in the high sound-speed simulation.

\clearpage
\begin{figure*}
\begin{center}
\plotone{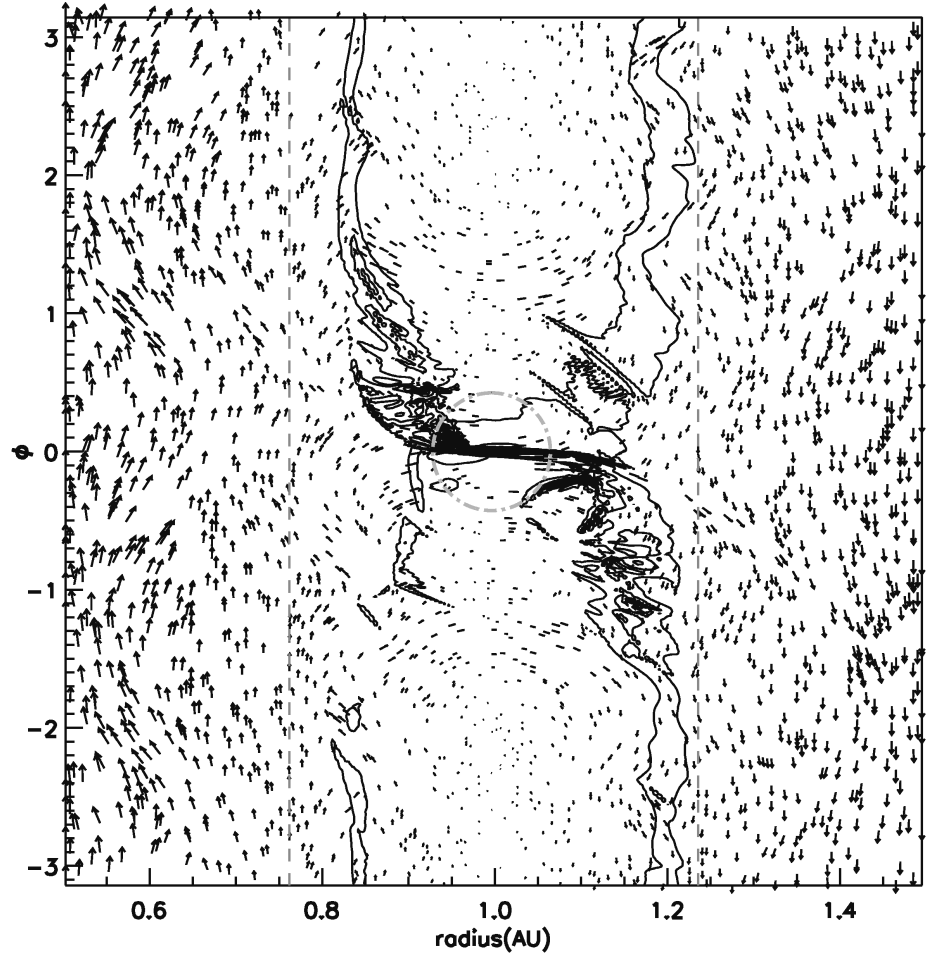}
\end{center}
\caption{Potential vorticity ($\varpi$) contours overlying the
velocity field of the disk for the low sound-speed simulation with
$\left(\frac{H}{a}\right) = 0.04$. Potential vorticity, conserved in
regions outside $R_R$, provides a good tracer of the streamlines
within the flow.}
\label{Fig:pv1}
\end{figure*}

\begin{figure*}
\begin{center}
\plotone{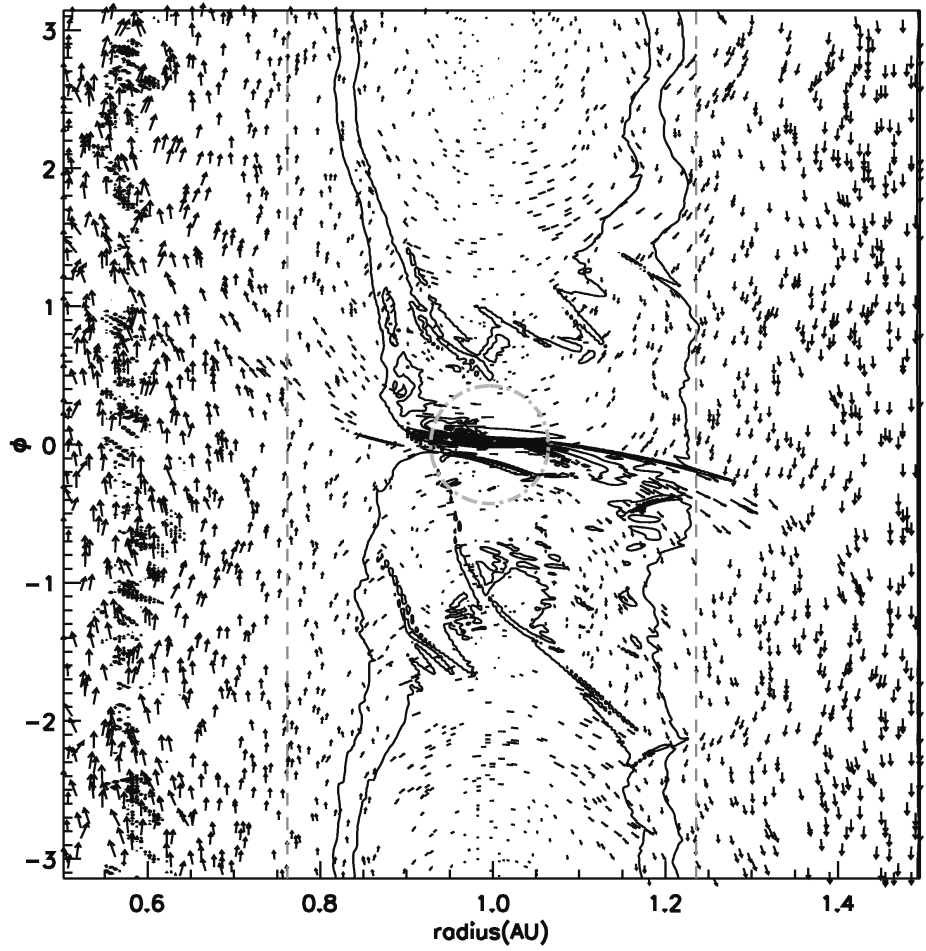}
\end{center}
\caption{The same as Figure (\ref{Fig:pv1}), but for
$\left(\frac{H}{a}\right) = 0.07$.}
\label{Fig:pv2}
\end{figure*}

\clearpage
In order to further illustrate that the pressure gradient is
insufficient to overcome the tidal potential barrier along the stream
line in the low sound-speed simulation, we identify streamlines with
contours of constant vortensity $\varpi$. Since shocks occur only
relatively close to the protoplanet, vortensity is approximately
conserved along the stream lines in regions outside the Roche lobe. In
Figures (\ref{Fig:pv1}) and (\ref{Fig:pv2}) we overlay the contours of
$\varpi$ on the velocity field. Outside $R_R$, the streamlines are
nearly parallel, and those that eventually enter the protoplanet's
Roche lobe are initially located at $\sim 2-3 R_R \sim 0.15-0.2 a$
from its orbit. 

\begin{figure*}
\begin{center}
\plotone{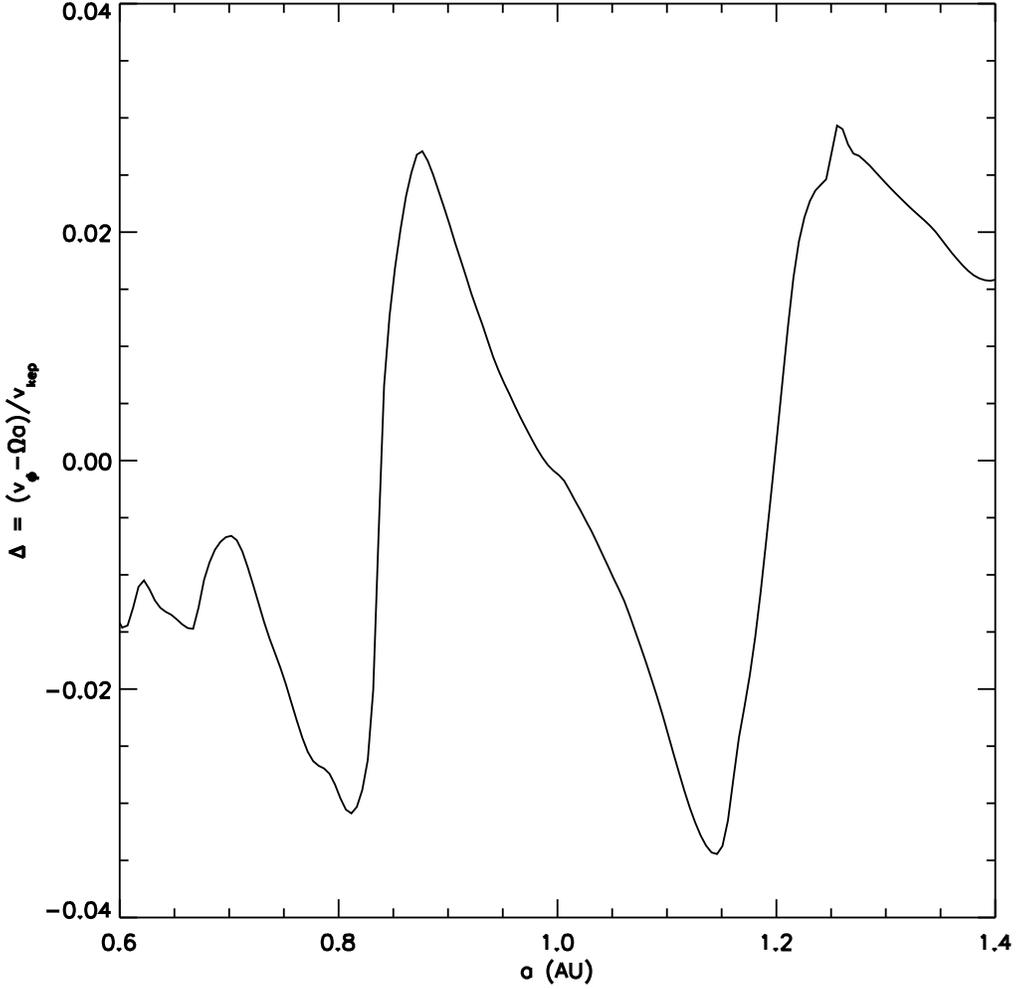}
\end{center}
\caption{ $\Delta =( v_y + \Omega a)/ v_{\rm kep} -1$ at the
protoplanet's superior conjunction for $\left(\frac{H}{a}\right) =
0.04$. This quantity parameterizes the deviation of the flow from
Keplerian ($\Delta=0$). $\Delta$ is related to $f_{\rm sub} \simeq 2
\left|\Delta\right|$, and can be used to determine the magnitude and
sign of $\beta$, the accretion rate parameter.}
\label{Fig:deltav1}
\end{figure*}

\begin{figure*}
\begin{center}
\plotone{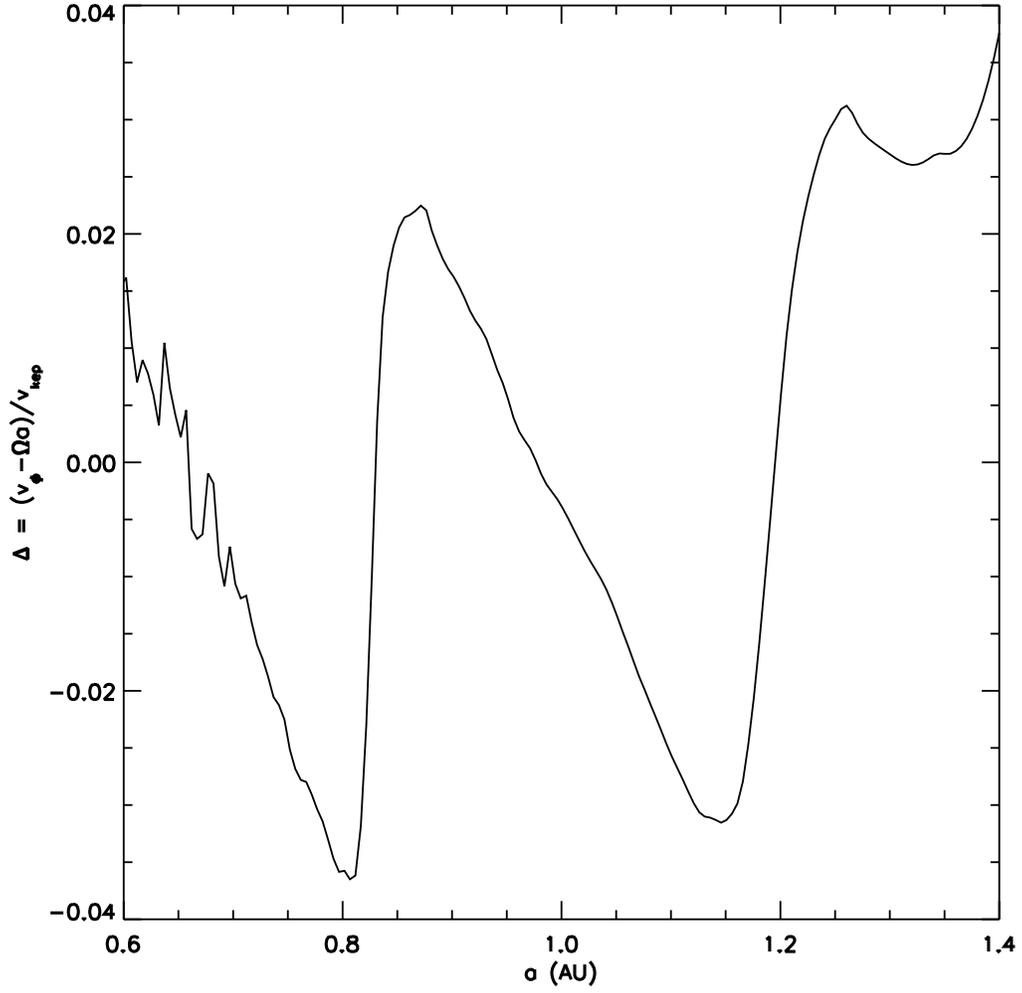}
\end{center}
\caption{Same as Figure (\ref{Fig:deltav1}) but for
$\left(\frac{H}{a}\right)= 0.07$.}
\label{Fig:deltav2}
\end{figure*}

Having determined that the flow is primarily along contours of
constant $\varpi$, we can utilize the discussion at the end of \S2.3
to link the flow velocity at the superior conjunction to the resulting
accretion rate. In Figures (\ref{Fig:deltav1}) and
(\ref{Fig:deltav2}), we plot $\Delta =\left(v_y + \Omega a\right)/
v_{\rm kep} -1$, where $v_{\rm kep} = (GM_\ast /a)^{1/2}$ and $v_y$ is
given in the rotating frame. Comparing to Equation (\ref{eq:fsub}), we
see that $f_{\rm sub} \simeq 2 \left|\Delta\right|$. At $a=0.8$AU and
$1.25$AU (which correspond to $\sim \pm x_{\rm max}$ away from the
protoplanet), $|\Delta |\sim 0.03$ and thus $f_{\rm sub}\sim
0.06$. For a streamline that connects $x_\infty = x_{\rm max}$ and
$x_s = R_R$, Equation (\ref{eq:beta2}) gives $\beta \sim 0.8-1.2$,
which is in agreement with the results shown in Figure
(\ref{Fig:beta}).

\clearpage

\begin{figure*}
\begin{center}
\includegraphics[height=4.0in,width=7.0in]{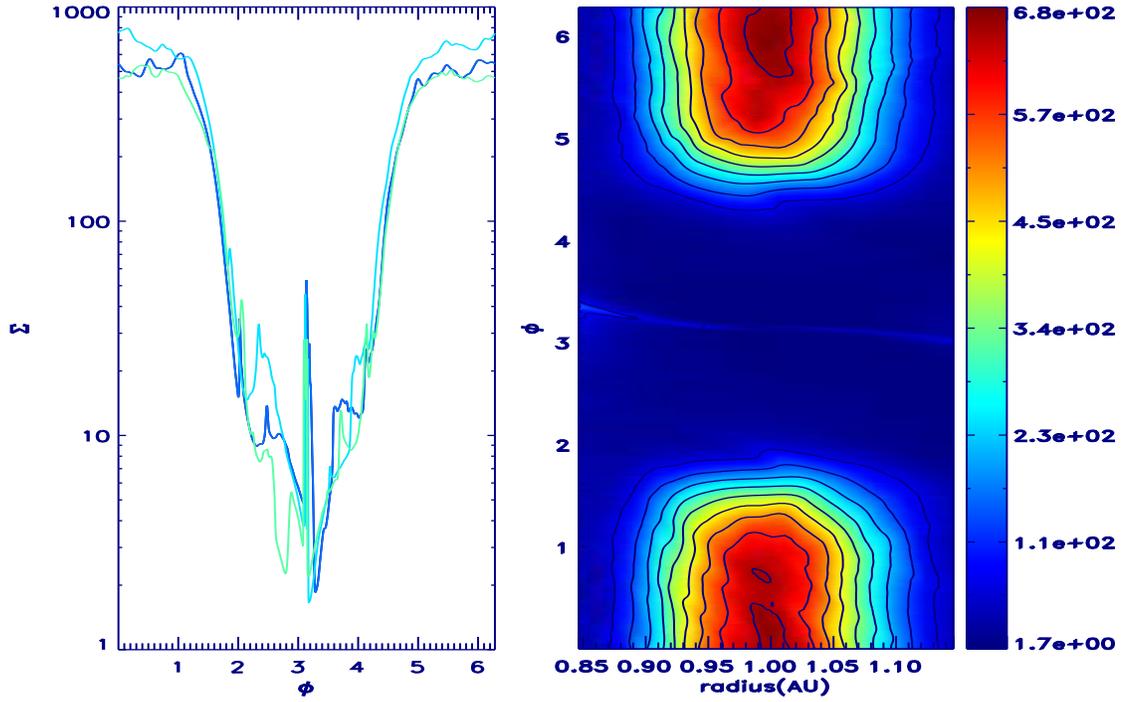}
\end{center}
\caption{Surface density ($\Sigma$) as a function of azimuthal angle
($\phi$) at several values of constant $r$ around the planet for
$\left(\frac{H}{a}\right) = 0.04$ at $t=500$ orbits (left
panel). Contours are at $r=0.95$, $1.0$, and $1.05$AU, while the
planet is located at $1$AU. The precipitous drop in density along
azimuth provides the necessary pressure gradient force needed to
overcome the tidal barrier. The right-hand panel shows the overall
surface density in the region.}
\label{Fig:den1} 
\end{figure*}

\begin{figure*}
\begin{center}
\plotone{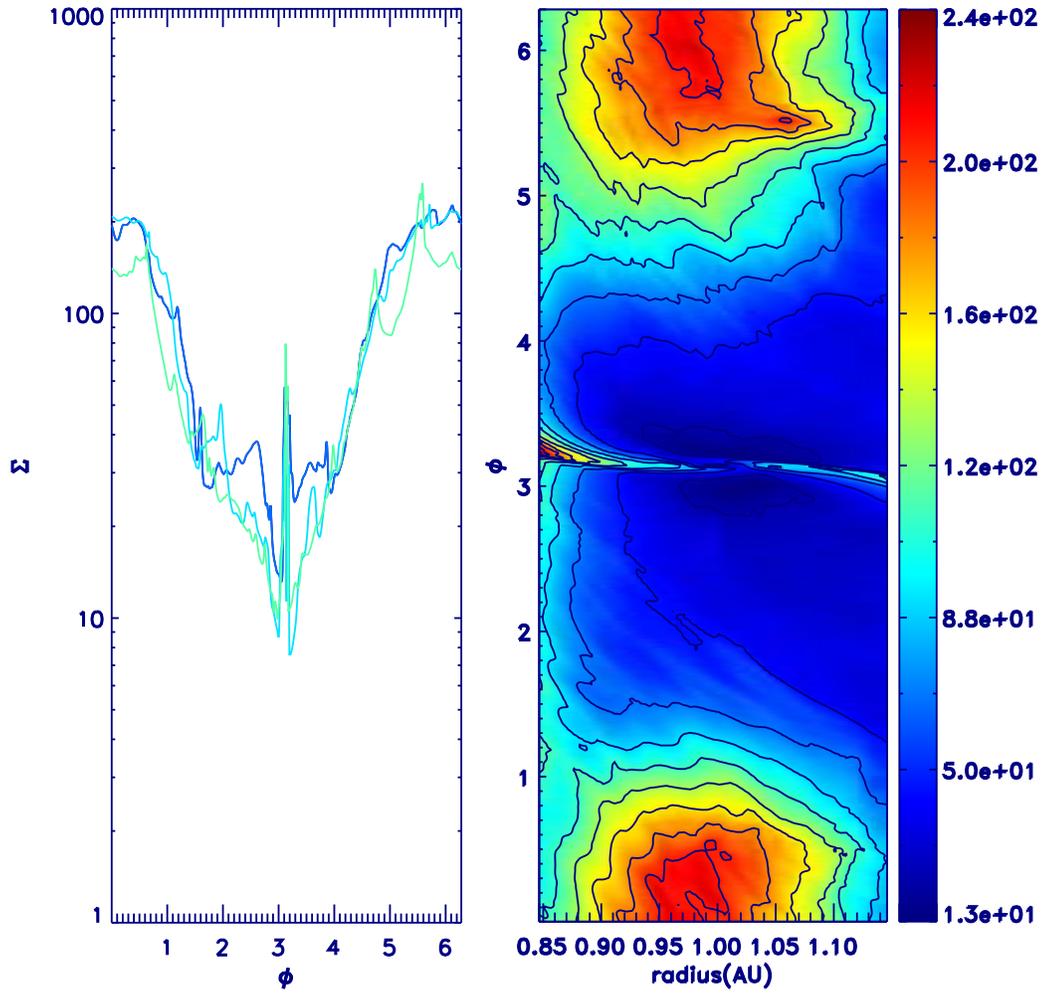}
\end{center}
\caption{Same as Figure (\ref{Fig:den1}) but for
$\left(\frac{H}{a}\right)= 0.07$. Given the higher sound speed, the
density drop needed to overcome the tidal barrier is not as large as
in the low sound-speed simulation.}
\label{Fig:den2}
\end{figure*}

\clearpage
In Figures (\ref{Fig:den1}) and (\ref{Fig:den2}), we plot the
variations of $\Sigma$ along the azimuth at constant semimajor
axis. Although the approximation of Equation (\ref{eq:2dbondi}) breaks
down near the protoplanet, it provides a reasonable indication of the
variation of $\Sigma$ along the streamlines. The results in Figures
(\ref{Fig:den1}) and (\ref{Fig:den2}) show much larger azimuthal
amplitude variations in $\Sigma$ for the low sound-speed model than
for the high sound-speed model. The magnitude of $\Sigma$ decreases
by two orders of magnitude in $H/a=0.04$ model. This result is
consistent with our analytic discussion that a large density gradient
is needed in the low sound-speed limit in order to sustain sufficient
pressure gradient to overcome the tidal potential barrier, and can be
compared to the 1D results of Figure (\ref{Fig:rho}).
\clearpage

\begin{figure*}
\begin{center}
\plotone{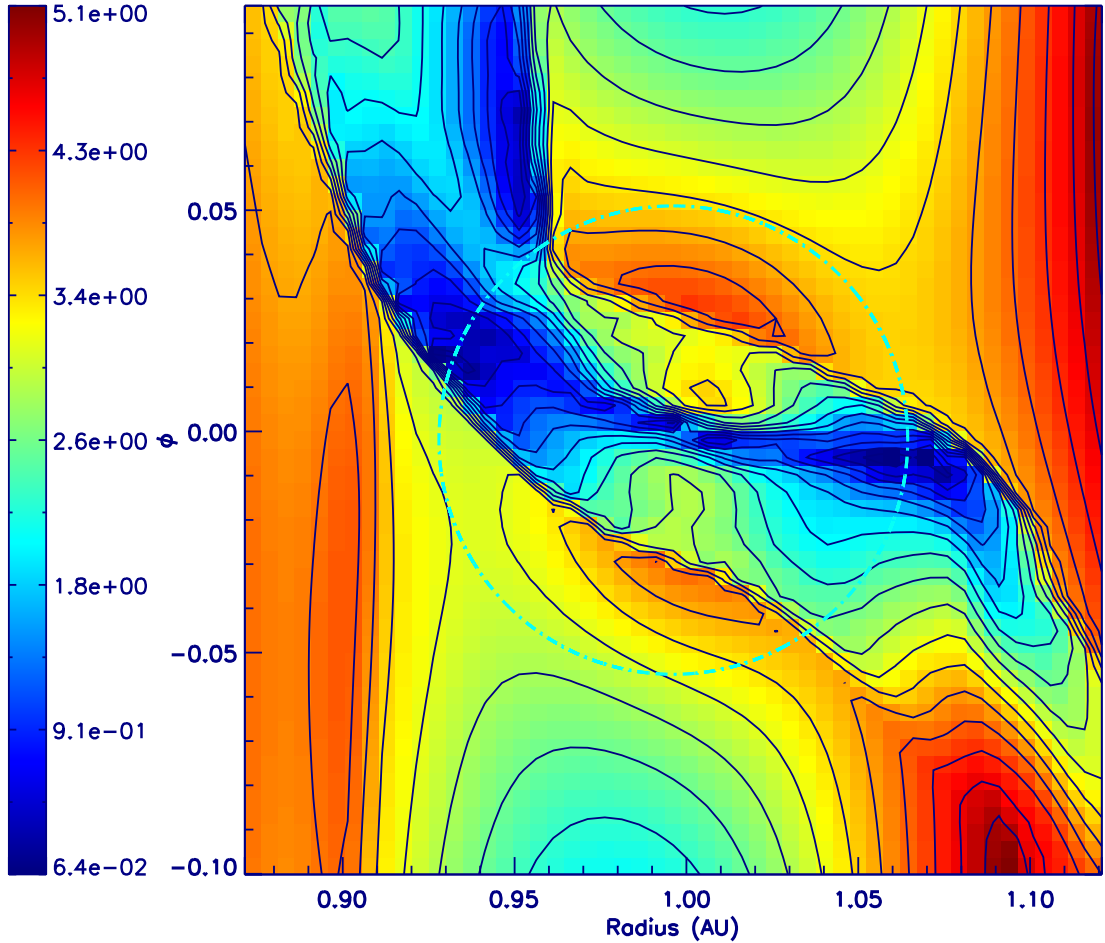}
\end{center}
\caption{The Mach number for $\left(\frac{H}{a}\right)=0.04$ in a
frame rotating with the planet. The dashed circle over-plotted on the
flow denotes the approximate Hill sphere. The subsonic region
surrounding the planet is well separated from the rest of the flow,
and a large density gradient is needed to overcome the tidal barrier.}
\label{Fig:v1}
\end{figure*}

\begin{figure*}
\begin{center}
\plotone{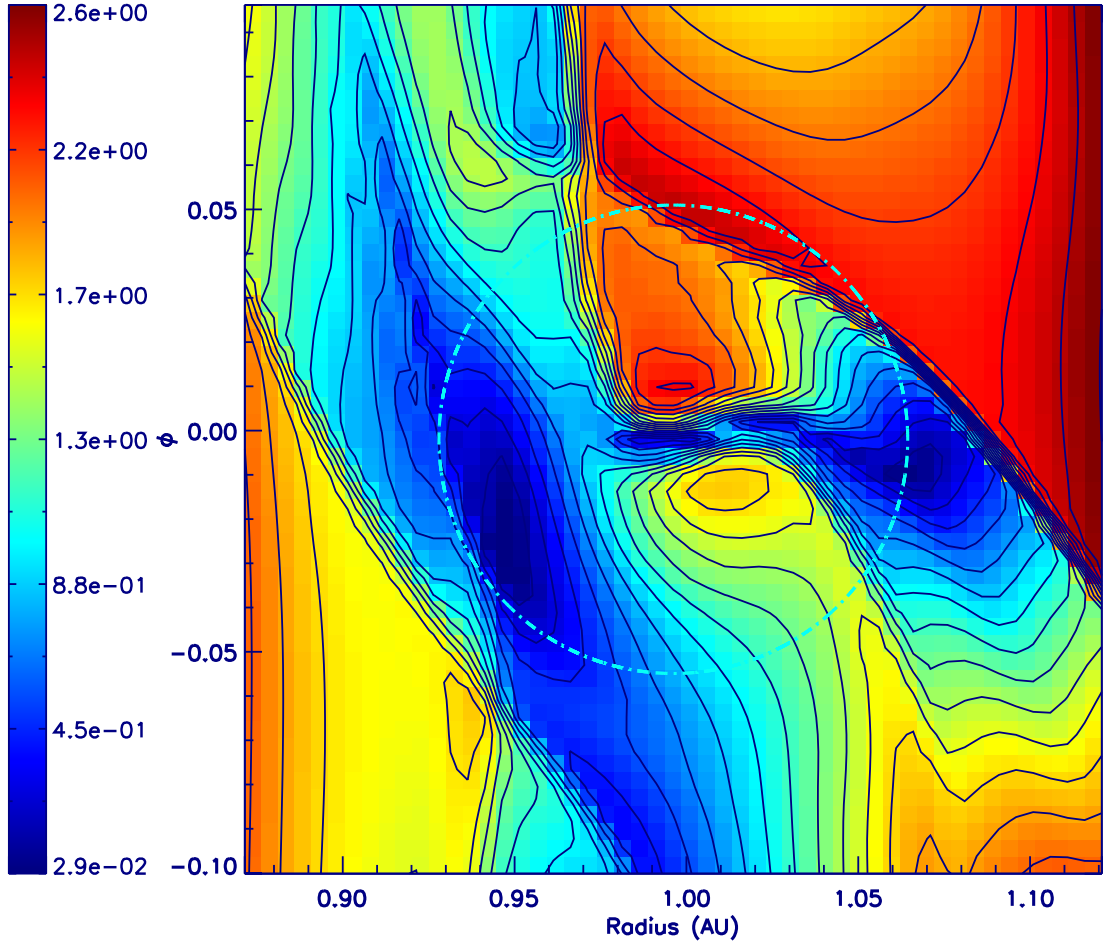}
\end{center}
\caption{The same as Figure (\ref{Fig:v1}) but for
$\left(\frac{H}{a}\right)=0.07$. The subsonic region surrounding the
planet is connected, via subsonic streamlines, to the material in the
feeding zone. Thus, the density gradient needed to overcome the tidal
barrier is not as large as in the $\left(\frac{H}{a}\right)=0.04$
simulation.}
\label{Fig:v2}
\end{figure*}

\clearpage
Figures (\ref{Fig:v1}) and (\ref{Fig:v2}) show the Mach number
($(v_x^2 + v_y^2)^{1/2}/c_s$) distribution in the rotating frame near
the planet for $(H/a)=0.04$ and $0.07$ respectively. Both the
approximately Keplerian flow far from the planet, and the horseshoe
orbits within the feeding zone are clearly visible in both plots. As
observed in \citet{Tanigawa2002}, the radial extent of the shocks from
the planet is proportional to the soundspeed. For the most part, the
subsonic region is bounded by a strong shock. However, for the high
sound speed case, the subsonic region also extends from inside the
Roche lobe to the feeding zone beyond $r_R$, allowing gas to
efficiently flow into the vicinity of the protoplanet. For the low
sound-speed model, The subsonic region near the protoplanet is well
separated from the low-velocity horse-shoe region. These figures once
again support the conjecture that large density gradient is needed for
the pressure to overcome the tidal potential barrier.
\clearpage

\clearpage

\subsection{Simulations Based on the 1D Tidal Barrier Approximation}
\label{sec:1dnumeric}
The main objective in our simulation is to determine the asymptotic
mass of protoplanets in depleting protostellar disks.  The results of
our analysis and numerical simulations (Figure (\ref{Fig:mplanet}))
indicates that protoplanets reach their asymptotic mass when their
growth timescale becomes comparable to that of the disk depletion
timescale, which is $10^5$ time longer than their orbital period.  The
accreted gas collapses to form the envelope of gas-giant planets with
size scales several times the present-day radius of Jupiter. The
dynamical timescale near the core is $\sim 10^{-5}$ that of the
orbital period. While we would like to simultaneously study the
evolution of the gaseous envelope and the decline of the mass influx
through the protoplanet's Roche lobe, the limitation of currently
available computational facilities prohibits the high-resolution,
multi-dimensional simulations of the planetary accretion process over
such a large range in the relevant timescales.

Here we carry out a series of 1D simulations based on the spherically
symmetric approximation of the tidal potential (\S\ref{sec:1dapprox}).
The basic equations for the 1D code include three conservation laws of
momentum, mass and energy (including radiative diffusion and
convection), and the equation of state. During the phase of dynamical
accretion, conventional numerical schemes based on a hydrostatic
approximation \citep{pollack1996} are no longer adequate for the
determination of the planetary structure and evolution. We developed a
numerical scheme which is based on a Lagrangian approach where the
radial coordinate evolves in time. It was developed to study the
envelope structure as it undergoes a transition from quasi hydrostatic
equilibrium to dynamical collapse (see a detailed description in
\citep{li2006}).  The advantage of this method is that we can use it
to resolve the large spatial range of the protoplanet's envelope and
to follow its evolution over an extended duration of time. The main
limitations of this method are the assumption of spherical symmetry
and our neglection of the protoplanet's tidal torque on the global
structure of the disk gas. We utilize the versatility of this scheme
to highlight the dominant physical effects and compare our results to
the 2D scheme to assess the validity of our 1D calculation at various
stages of protoplanetary growth.

For the present application, gas is accreted onto a Saturn-mass object
($M_p = 100 M_\oplus$).  During the dynamic gas accretion stage of
planetary formation, the envelope of the proto-planet will contract
much more rapidly than before. While the gas flow close to the core
remains in quasi-equilibrium, the outer region of the envelope,
especially the region near the surface of the proto-planet, should
experience dynamic gas accretion. We focused our calculations on the
dynamic region of the gas flow, which covers the radial region from 10
core radii to 2 Bondi radii. Across the inner boundary, we calculated
the gas accretion rate and infer a luminosity based on the assumption
that the inflow is halted at the core radius.  As the gas flow in this
region tends to a steady-state (to be shown below), the radiative
transfer is very efficient and there is insufficient thermal pressure
to counter balance the core's gravity throughout this region.  It is
therefore a reasonable assumption for the velocity across the inner
boundary to be close to the free-fall velocity. Constant temperature
and density are set at the outer boundary to denote the nebula
conditions. In order to approximate the effects of the tidal
potential, we added a simple prescribed outward gravitational force in
the equation of motion (Equation \ref{eq:1dpot}). We can then adjust
the location of the Roche radius where gravity vanishes in order to
assess the effects of this tidal barrier.

For comparison, we first constructed a test model for standard Bondi
accretion with the nebula conditions of $T_{neb}\sim 50K$ and $\rho
_{neb}\sim 1\times 10^{-14} g/cm^3$ (Model $0$ in Table \ref{tbl1}).
These parameters are relevant in the outer ($>5-10$ AU) regions after
a substantial fraction ($\sim 10^{-3}-10^{-2}$) of the minimum mass
nebula has already been depleted through global depletion or gap
formation. In this, we neglect the radiative energy transfer and tidal
barrier effects so that we can compare our numerical results with the
the analytic Bondi solution. We find a gas accretion rate of around
$3\times 10^{17} g/s$ from our numerical calculations, which is in
agreement with the analytic Bondi solution $\dot{M}\simeq
1.4\times10^{11}({M\over {M_\odot}})^2({\rho(\infty) \over
{10^{-24}}})({c_s(\infty)\over {10km s^{-1}}})^{-3} g/s$
\citep{Bookaccretion}.

\begin{figure*}
\begin{center}
\plotone{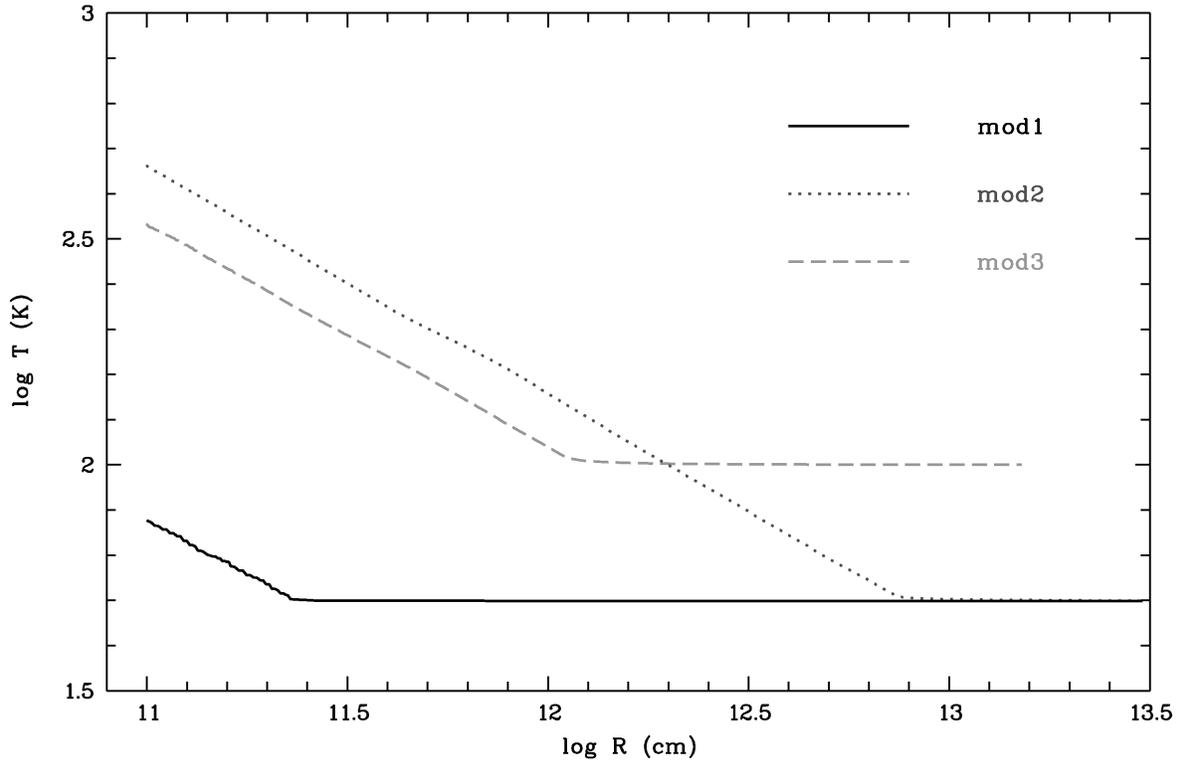}
\end{center}
\caption{Temperature profiles as a function of radius for 1D models 1,
2, and 3 after they have reached steady-state. The parameters for each
model are listed in Table (\ref{tbl1}).}
\label{Fig:temp}
\end{figure*}

\begin{figure*}
\begin{center}
\plotone{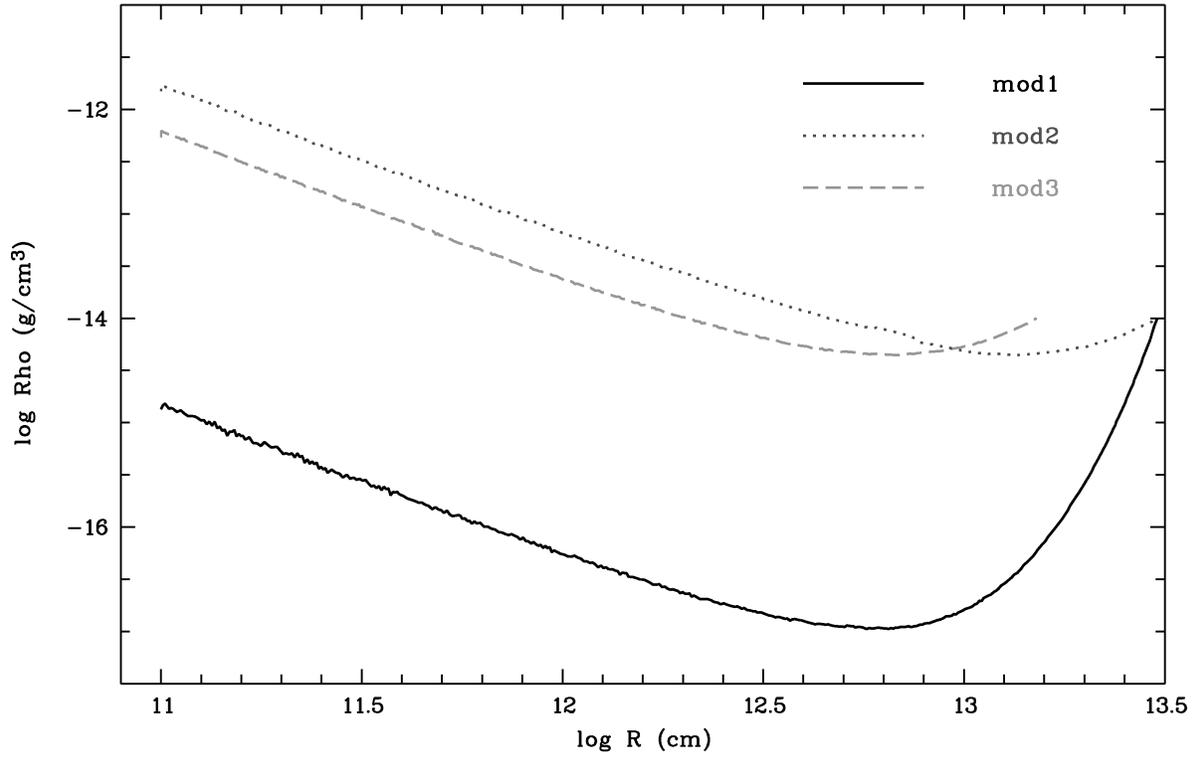}
\end{center}
\caption{Density profiles as a function of radius for the three 1D models
after they have reached steady-state. The denotations are the same as
in Figure (\ref{Fig:temp}).}
\label{Fig:rho}
\end{figure*}

\begin{figure*}
\begin{center}
\plotone{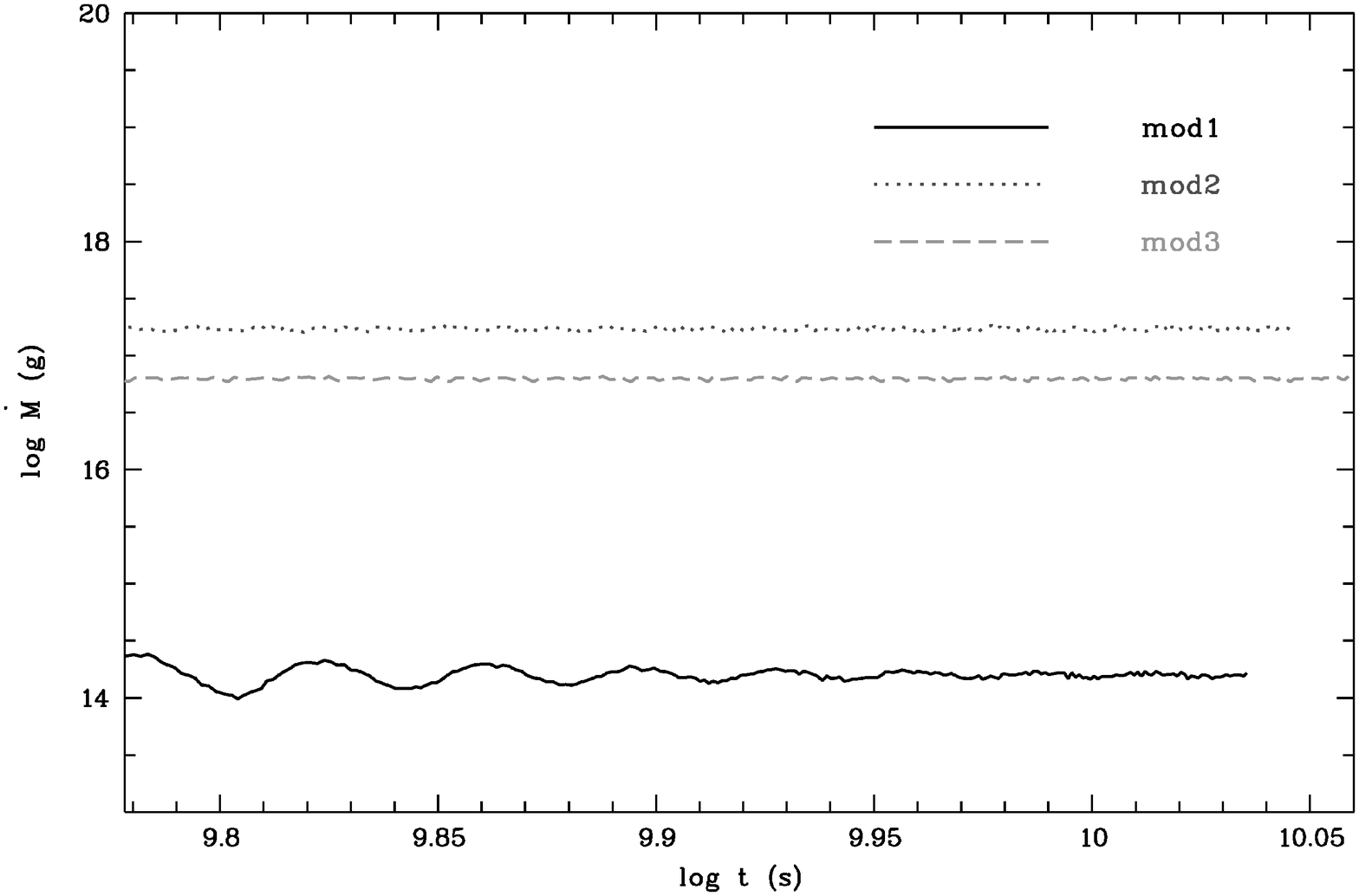}
\end{center}
\caption{Gas accretion rate as function of time for the three models. 
Different line types correspond to the different models listed in 
Table (\ref{tbl1}).}
\label{Fig:mdot}
\end{figure*}

We constructed another three models to investigate the tidal barrier
effects, with different parameters listed in Table \ref{tbl1} (Model
$1, 2, 3$). In the table, $T_{neb}$ denotes the temperature of the
nebula gas, which corresponds to the sound-speed at the outer boundary
of the envelope. The third column shows the ratio of Roche radius to
the Bondi radius. The last column lists the average gas accretion
rate for each model. The outer boundary condition for density is
$1\times 10^{-14} g/cm^3$ for all the models.
\clearpage
\begin{table}
\begin{center}
\begin{tabular}{@{}|c c c c @{}|}
\hline ${\rm Model\ No.}$ & ${\rm T_{\rm neb}}\,({\rm K})$ & ${\rm
   R_{R}/R_{B}}$ & ${\rm \stackrel{.}{M}\,({\rm g/s})}$ \\ \hline
   ${0}$ & 50 & - & ${\rm 3.0\times 10^{17}}$ \\ 1 & 50 & 0.5 &
   ${\rm 1.6\times 10^{14}}$ \\ 2 & 50 & 1.0 & ${\rm 1.7\times
   10^{17}}$ \\ 3 & 100 & 1.0 & ${\rm 6.2\times 10^{16}}$ \\ \hline
\end{tabular}
\caption{Parameters for the single, one-dimensional models. Model 0
shows Bondi accretion, which does not include radiative energy
transfer or the tidal barrier. $T_{\rm neb}$ denotes the temperature
of the nebula gas, while $\dot{M}$ is the average gas accretion rate
calculated from the model.\label{tbl1}}
\end{center}
\end{table}
\clearpage
We evolved all the models to steady-state, where the accretion rate
became constant with radius. Figure (\ref{Fig:temp}) shows the
temperature profiles of the three models after they have reached
steady-state. As we can see from the figure, the outer regions of the
envelopes are optically thin, cool, and isothermal for all the models.
Figure (\ref{Fig:rho}) plots the density profiles of the three
models. The model with $R_R/R_B\sim 0.5$ shows a much larger density
gradient and thus much smaller density at the location of Roche radius
than the other two models. At small radii, the pressure gradient does
not significantly perturb the nearly free-falling motion of the
gas. However, as the stellar gravity acts to repel the gas outside the
Roche radius, sufficiently large positive pressure and density
gradients are needed to overcome the tidal barrier. Therefore, in the
case of $R_B > R_R$, this gradient leads to a density at $R_R$ that is
much smaller than that in the disk. Since free fall onto the core can
only proceed inside $R_R$ and the flow across $R_R$ is limited by the
speed of sound, the mass accretion rate is much reduced from the
conventional Bondi formula. Figure (\ref{Fig:mdot}) shows the
evolution of gas accretion rate with time at the locations of Roche
radius for three models. Nearly constant accretion rate indicates that
all the models have reached steady state after approximately $6\times
10^9 s$. Since the flow between $R_B$ and $R_R$ is significantly
perturbed by the gas pressure gradient and softened gravity, this time
scale must be at least several times the local dynamical free-fall
time scale ($\sim 5-8 \times 10^8$s). However, in contrast to the 2D
simulations, the flow does not need to adjust over the liberation
period of flows in the horseshoe region and the synodic period of the
flow near the separatrix at the Lagrangian points.

From the two models with the same $R_R/R_B$ but different nebula
temperature (Model 2 and Model 3), it's found that higher nebula
temperature leads to slightly lower gas accretion rate. This behavior
is similar to what we find in the Bondi solution, in which
$\stackrel{.}{M} \propto c_s^{-3}(\infty)$. When comparing the models
with the same nebula temperature but different $R_R/R_B$ (Model 1 and
Model 2), we find that the accretion rate is reduced by three orders
of magnitudes for the model with $R_B > R_R$. Thus, the tidal barrier
significantly suppresses in the gas accretion across the Roche
radius. Even if the models are surrounded by nebula gas with the same
density, only a small fraction of that gas can be accreted by the
model with $R_B > R_R$.
\clearpage

\clearpage

\begin{figure*}
\begin{center}
\includegraphics[scale=0.50,angle=-90]{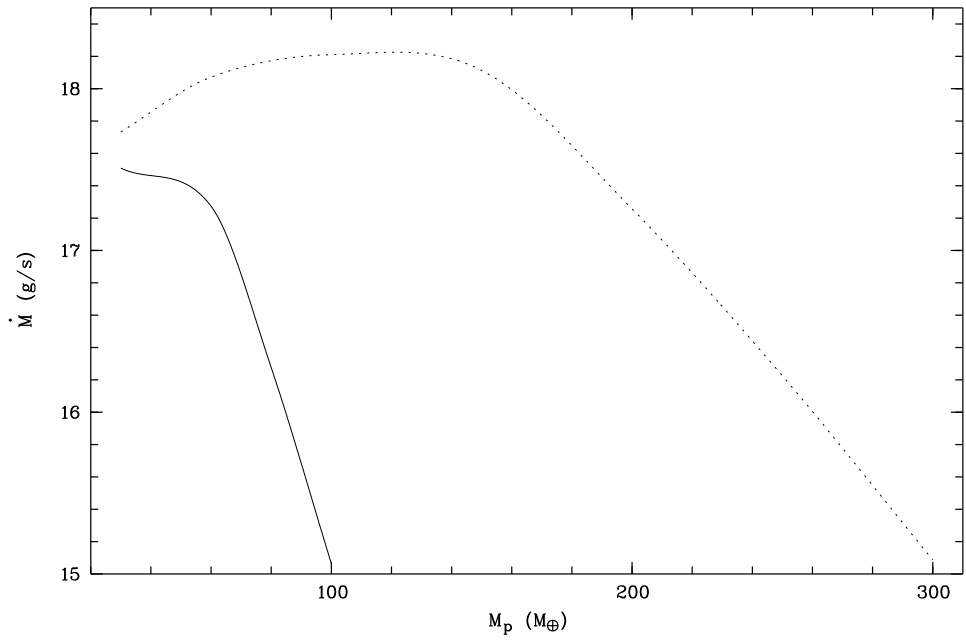}
\end{center}
\caption{The steady-state gas accretion rate as a function of $M_p$
for models series 4 and 5 are represented by solid and dotted lines
respectively. Both simulations show a significant decrease in
$\dot{M}$, but it occurs at much lower mass for the low temperature
nebula.}
\label{Fig:mdotofmp}
\end{figure*}

\begin{figure*}
\begin{center}
\includegraphics[scale=0.50,angle=-90]{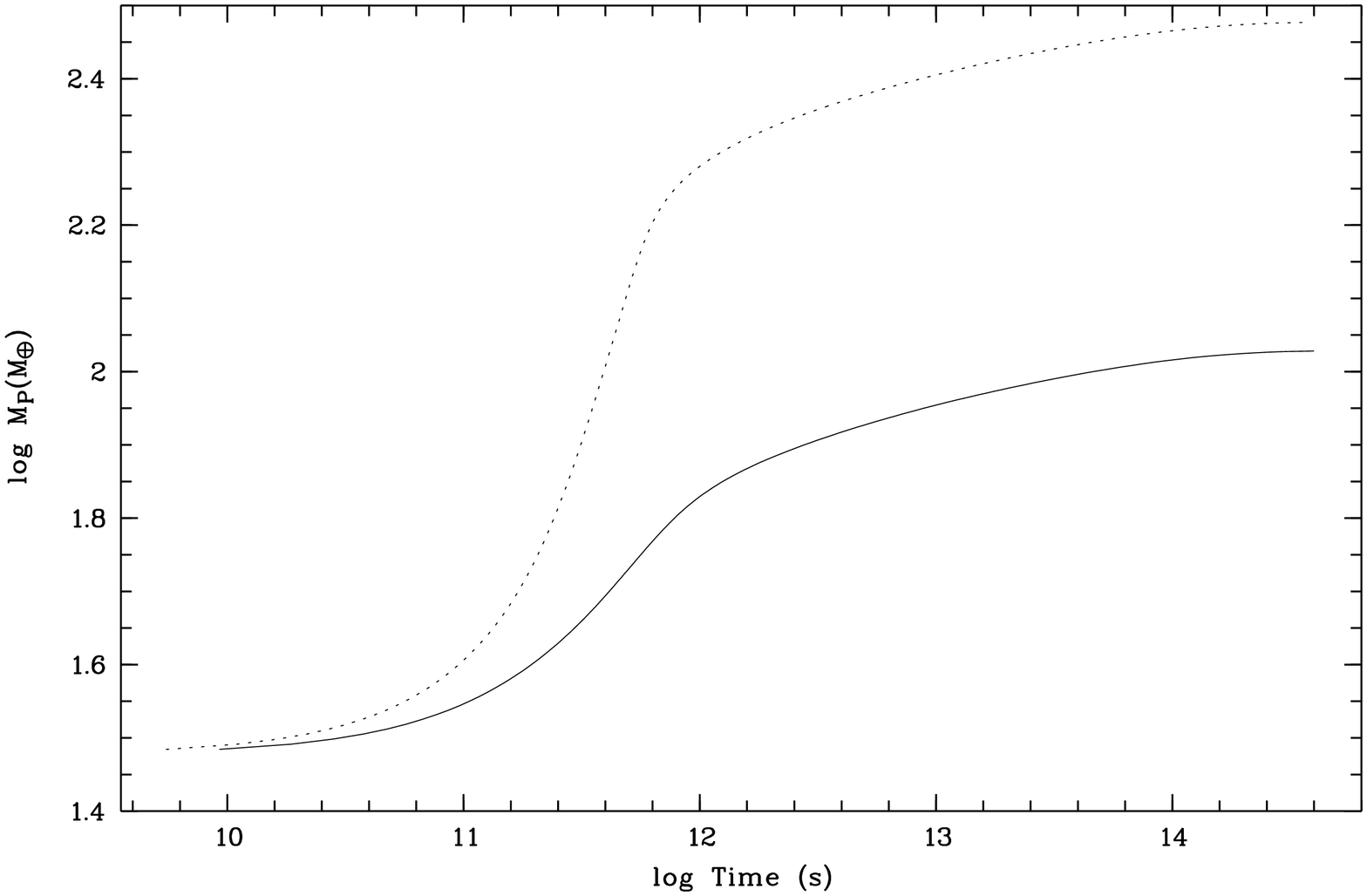}
\end{center}
\caption{The planet's mass after 20 Myr of gas accretion, starting
with $M_p = 30 M_\oplus$ initially.  The gas depletion time scale is
set to be $\tau_{\rm dep}=3$ Myr. The solid line designates the model
series with $T_{neb}=50K$, while the dotted line is for the model
series with $T_{neb}= 100K$. As suggested from Figure
\ref{Fig:mdotofmp}, and seen in Figure \ref{Fig:mplanet}, the
asymptotic mass of the lower temperature model series is significantly
lower then in the high $T_{neb}$ model series.}
\label{Fig:mp1e6}

\end{figure*}

We further compute two series of disk models (4 and 5) for planetary
masses ranging from $30M_{\oplus}$ to $100$ and $300 M_{\oplus}$,
respectively. Models in series $4$ have $(T,\rho_{neb})$ of $(100 {\rm
K}, 10^{-12} {\rm g \ cm^{-3}})$, while models in series $5$ have
$(T,\rho_{neb})$ of $(50 {\rm K}, 10^{-12} {\rm g \ cm^{-3}})$. We
determine the magnitude of $\dot M$, as a function of $M_p$ after it
has attained a steady state value at 10 AU (see Figure
\ref{Fig:mdotofmp}). These results are in good agreement with the
relationship in Equation (\ref{eq:1dmdot}).  Based on these results,
we determine the total mass of the protoplanet after 20 Myr of
accretion onto a core with an initial mass of $30 M_\oplus$ (see
Figure \ref{Fig:mp1e6}). In both model series we neglect the decline
of gas density associated with formation of the gap in the disk, but
include that due to the global depletion over a time scale $\tau_{\rm
dep} = 3 $Myr.  Despite these simplifications, the asymptotic values
of $M_p$ (100 and 300 $M_\oplus$ for $T_{neb}=50$K and $100$K
respectively) are comparable to that found in Equation
(\ref{eq:mfinal1}) with the appropriate boundary conditions.

\section{Summary and Discussion}

In this paper, we present evidence to suggest that the asymptotic mass
of protoplanets is determined by the structure of their nascent disk.
The main physical process involved here is that the tidal potential of
the central star provides a barrier to the gas flow in the vicinity of
the protoplanet. Through its tidal interaction with the disk, a
growing protoplanet exerts an increasingly strong torque which first
leads to the formation of a gap. Although the disk gas may continue to
diffuse into the gap, it is forced to flow along the horseshoe stream
lines. In the absence of any dissipation, the vortensity of the flow
along the stream lines is conserved. But near the protoplanets, shock
dissipation can lead to vortensity generation and dissipation.

In the limit that the protoplanets' mass is small and its Roche radius
is smaller than the disk thickness, the tidal potential barrier is
shallow and modest surface density variation would provide adequate
pressure gradient to overcome the tidal potential barrier while
preserving vortensity along the stream line. However, in cold disks
with thicknesses less than the protoplanets' Roche radius, a very
large surface density gradient is required to create the critical
pressure gradient force needed to overcome the tidal barrier along the
stream line. Consequently, mass supply from the feeding zone into the
vicinity of the protoplanet is quenched. In a pressure-free gas disk,
the horseshoe stream lines of cold gas cannot join onto those in the
disk around the protoplanet. Only particles with a limited range of
initial orbital parameters that do not follow streamlines can directly
strike the protoplanet.

Based on this consideration, we suggest that the asymptotic mass of
the protoplanet is determined by the condition $R_R \sim H$. This
condition is similar to the thermal criterion for gap formation
\citep{lin1986}. When the planet approaches this value, the mass supply
into the gap is much reduced and the dynamics of the gas is slightly
altered \citep{bryden1999}. Although these consequence do not directly
terminate protoplanets' growth, they contribute to the reduction of
the accretion rate onto the protoplanet.

The gaseous material which enters into the protoplanet's Roche lobe
forms a circumplanetary disk. Angular momentum is transferred from the
circumplanetary disk by both tidally induced spiral shocks and
turbulent viscosity. Initially, the flux through the circumplanetary disk
is high and the disk environment is prohibitive for the formation of
regular satellite systems \citep{lunine1982,canup2002,mosqueira2003_1,
canup2006}. However, our results indicate that during the final stage
of the protoplanet's growth, the characteristic timescale for
protoplanetary accretion increases beyond the circumstellar disk
depletion timescale. It is at this stage that physical environment in
the circumplanetary disks become less hostile to satellite
formation \citep{canup2006}. The studies of satellite formation process is
beyond the scope of the present paper and they will be presented
elsewhere.

In the final analysis, we are interested in the origin of the
present-day mass distribution of extrasolar planets which, is observed
to be $dN/dM \propto M_p ^{-1}$ to $M_p^{-1.4}$ in the range of a
fraction to several $M_J$ \citep{marcy2005}. This distribution also
has a cut-off below $M_p \sim 10-20 M_J$. Since the host stars of most
known extrasolar planets have masses $M_\ast \sim M_\odot$, the
corresponding $q= M_p/M_\ast$ has a similar distribution $d N / d q
\sim q^{-1}$ to $q^{-1.4}$ in the range $\sim 10^{-3}$.

Scaling the formula derived in \S\ref{sec:loading} to the values used
in this simulation and assuming a dissipation rate of $\tau_{dep} =
3-5\mathrm{Myr}$ and $\beta = 1$, $M_p$ is comparable to the mass of
Saturn for an aspect ratio of $0.04$ and that of Jupiter for an aspect
ratio of $0.07$. The most favorable location for the formation of gas
giant planets is near the snow line $a_{\rm ice}$ \citep{ida2004_1}.
The magnitude of $a_{\rm ice}$ is determined by both the gas accretion
rate within the disk and the intensity of stellar irradiation and
during the classical T-Tauri phase, and $a_{\rm ice}$ can evolve over
nearly an order of magnitude in scale \citep{garaud2006}. The disk
aspect ratio at the snow line $(H\left(a_{\rm ice}\right)/ a _{\rm
ice})$ can also change by a factor of two. Depending on the epoch of
gas-giant planet formation, the variation in $(H\left(a_{\rm
ice}\right)/a _{\rm ice})$ can lead to an order of magnitude spread in
the asymptotic mass of the planets. Thus, the observed planetary mass
distribution may be used as to infer the ratio of the planet formation
and disk depletion timescales. Detailed simulations of such models
will be presented elsewhere.

Target lists for planet searches using the radial velocity technique
have recently been expanded to include M-dwarf stars. However, the
detection frequency for Jupiter mass planets around these low-mass
stars appears to be much lower than that around G dwarfs. Until
recently, the only known member of this population with known
Jupiter-mass planets was GJ 876. The lack of massive gas-giants is
also conspicuous in the microlensing survey for extrasolar planets
\citep{beaulieu2006}. The difficulty for gas-giant planet formation
may be due to the inadequate amount of heavy elements for assembling
sufficiently massive protoplanetary cores \citep{ida2004_1} or
inability to overcome type I migration and retain sufficiently mass
cores to initiate the process of rapid gas
accretion\citep{ida2006}. In this paper, we show that even in the
low-probability event that proto gas-giant planets may form, their
asymptotic mass may severely limited by the tidal barrier in
accordance with Equation (\ref{eq:mfinal1}).  If the surface densities
of both gas and dust increases with $M_\star^2$ as suggested by
observationally inferred $\dot M$'s from the disks onto their central
stars, the onset of dynamical gas accretion would occur at much later
epoch with a relatively small total gas content around M dwarfs than
around solar type stars.  At this stage, the value of $H(a_{\rm
ice})/a$ is also an slowly increasing function of $M_\ast$.  All of
these effects combine sensitively to quench the gas accretion rate and
severely limit the asymptotic mass of the planets well below that of
Jupiter.

The co-existence of multiple gas-giant planets around the same host
star suggests that their asymptotic mass is limited by local tidal
truncation (through gap formation) rather than the global depletion of
the disk. The asymptotic mass distribution within such systems is
determined by both the planets' formation location and epoch. In a
steady-state disk, the value of the aspect ratio $H/a$ determined from
the disk mid-plane structure generally increases with $a$
\citep{garaud2006}. In a viscous evolving disk, $\Sigma$ in the outer
regions of the disk decreases with $a$ much more rapidly than in a
steady disk. In general, $H/a$ attains a maximum at a radial location
$a_{\rm max}$ which is an increasing function of the the accretion
rate through the disk. During the depletion of the disk, $a_{\rm max}$
decreases. Outside $a_{\rm max}$, the disk is shielded from the
stellar irradiation and the density scale decline rapidly with radius.
This process may determine the mass distribution within multiple gas
giant planet systems. It may also limit the domain of gas-giant
formation.

\acknowledgements We thank Dr. P. Bodenheimer, S. Ida, and Hui Li for
useful conversation and Dr. Geoff Bryden for making available a 2D
numerical hydrodynamic scheme. This work was supported in part by
NASA (NAGS5-11779, NNG06-GF45G, NNG04G-191G, NNG05-G142G), JPL
(1270927), NSF (AST-0507424), IGPP, and the California Space
Institute.

\bibliographystyle{aa}
\bibliography{ian}

\end{document}